\documentclass[10pt,twocolumn,twoside]{IEEEtran}
\usepackage{graphicx}
\usepackage{epsfig}
\usepackage{amssymb}
\usepackage{enumerate}
\usepackage{latexsym}
\usepackage[cmex10]{amsmath}
\newtheorem{definition}{Definition}

\newtheorem{remark}{Remark}

\newtheorem{lemma}{Lemma}
\def\dif{ {\rm d} }
\def\deltt{\Delta_{\rm t}} 
\def\fN{f^{}_{\cal N}}
\def\EqualDist{\mathrel{\mathop=^{\rm d}}}
\def\EqualDef{\mathrel{\mathop=^{\rm def}}}

\def\bSigma{{\boldsymbol\Sigma}}

\def\bxi{{\boldsymbol\xi}}
\def\beeta{{\boldsymbol\eta}}

\def\bZ{{\boldsymbol Z}}

\def\bF{{\boldsymbol F}}

\def\bz{{\boldsymbol z}}
\def\bX{{\boldsymbol X}}

\def\bY{{\boldsymbol Y}}

\def\mbZ{{\boldsymbol Z}}

\def\mbu{{\boldsymbol U}} 

\def\mbv{{\boldsymbol V}} 

\def\bQ{{\boldsymbol Q}}
\def\bA{{\boldsymbol  A}}

\def\bC{{\boldsymbol  C}}

\def\bB{{\boldsymbol  B}}

\def\bG{{\boldsymbol  G}}

\def\bH{{\boldsymbol H}}
\def\bI{{\boldsymbol  I}}
\def\bK{{\boldsymbol K}}

\def\bL{{\boldsymbol  L}}

\def\bZ{{\boldsymbol  Z}}
\def\bP{{\boldsymbol  P}}

\def\bR{{\boldsymbol  R}}
\def\br{{\boldsymbol  r}}
\def\bu{{\boldsymbol  U}}

\def\bS{{\boldsymbol  {S}}}
\def\bs{{\boldsymbol  s}}
\def\bT{{\boldsymbol  T}}

\def\bJ{{\boldsymbol  J}}

\def\bv{{\boldsymbol V}}

\def\cov{ {\rm cov} }
\def\rel{ {\rm rel} }
\def\tr{{\rm tr}}
\def\ri{ {\rm i} }
\def\re{ {\rm e} }

\def\eye{{\rm i}}
\def\deltt{\Delta_{\rm t}}
\def\e{{\rm e}}
\def\EqualDist{\mathrel{\mathop=^{\rm d}}}

\def\sxy*{\mathbf{S}_{xy^{*}}(f)}
\def\syx*{\mathbf{S}_{y^{*}x}(f)}

\begin{document}

%
%
\title{A Frequency Domain Test for Propriety of Complex-Valued Vector Time Series}

\author{Swati~Chandna
        and~Andrew~T.~Walden,~\IEEEmembership{Senior~Member,~IEEE}
\thanks{
Copyright (c) 2016 IEEE. Personal use of this material is permitted. However, permission to use this material for
any other purposes must be obtained from the IEEE by sending a request to pubs-permissions@ieee.org. 
S.~Chandna is with Dept. of Statistical Science, University College London, London WC1E 6BT, UK
 (e-mail: s.chandna@ucl.ac.uk). A.~T.~Walden is with the Dept. of Mathematics, Imperial College London, London SW7 2AZ, UK (e-mail: a.walden@imperial.ac.uk).} }
\IEEEpubid{}
\maketitle
\begin{abstract}
This paper proposes a frequency domain approach to test the hypothesis that a complex-valued vector time series is proper, i.e., for testing whether the vector time series is uncorrelated with its complex conjugate. If the hypothesis is rejected, frequency bands causing the rejection  will be identified and might usefully be related to known properties of the physical processes. The test needs the associated spectral matrix which can be estimated by multitaper methods using, say, $K$ tapers. Standard asymptotic distributions for the test statistic are of no use since they would require $K \rightarrow \infty,$
but,
as  $K$ increases so does resolution bandwidth which causes spectral blurring. In many analyses $K$ is necessarily  kept small, and hence our efforts are directed at practical and accurate methodology for hypothesis testing for small $K.$  Our generalized likelihood ratio statistic combined with exact cumulant matching 
gives very accurate rejection percentages and outperforms other methods. We also  prove that the  
statistic on which the test is based is comprised of canonical coherencies arising from our complex-valued vector time series.
Our methodology is demonstrated on ocean current data collected at different depths in the Labrador Sea.

Overall this work  extends results on propriety testing for complex-valued vectors to the complex-valued vector time series setting.
\end{abstract}

%
%
%

\section{Introduction}
There has long been an interest in time series motions on the complex plane: the rotary analysis method decomposes such motions into counter-rotating components  which have proved particularly useful in the study of geophysical flows influenced by the rotation of the Earth \cite{ElipotLumpkin08,EmeryThomson98,Mooers73,vanHarenMillot04,Walden13}. 

Let a complex-valued  $p$-vector-valued
discrete time series be denoted $\{ {\bZ}_t \}.$ This has as
$t$-th element, ($t \in \mathbb{Z}),$ the column vector 
${\bZ}_t=[Z_{1,t},\ldots,Z_{p,t}]^T.$ A length-$N$ realization of $\{\bZ_t\}$ namely $\bz_0, \ldots, \bz_{N-1}$  has $\bz_t \in \mathbb{C}^p.$ In this paper we assume the $p$ processes are jointly second-order stationary.

We propose a frequency domain approach to testing the hypothesis that a complex-valued $p$-vector-valued time series is proper, i.e., for testing whether the vector time series  $\{\bZ_t\}$ is uncorrelated with its complex conjugate $\{ \bZ^*_t\}.$ If we denote the covariance sequence  between these terms by $\{\br_{\bZ,\tau}\}$ then propriety corresponds to  $\br_{\bZ,\tau}={\bf 0}\,\,\text{for all}\, \tau\in {\mathbb Z},$ or  ${\bR}_{\bZ}(f)={\bf 0}$ over the Nyquist frequency range, where ${\bR}_{\bZ}(f)$ is the Fourier transform of $\{\br_{\bZ,\tau}\}$. Otherwise the time series is said to be improper;
the practical importance and occurrence of improper processes is discussed in, e.g., \cite{Adali_etal11}, \cite{Navarro-Moreno_etal09} and \cite{SchreierScharf10}.

The relevance of propriety for 
two-component complex-valued series ($p=2)$ can be found in \cite{Mooers73}. Because the series are complex, two types of cross-covariance can be defined: that  between the two series, known as the inner cross-covariance \cite{Mooers73}, and that beween one series and the complex conjugate of the other, known as the outer cross-covariance \cite{Mooers73}. If the vector time series is proper then 
the outer cross-covariance is everywhere zero.

In this paper we take as an example a six-component complex-valued ocean current time series recorded in the Labrador Sea.
Frequency domain analysis is particularly useful in a scientific setting: if the hypothesis is rejected, frequency bands causing the rejection can be identified and quite possibly related to known properties of the physical processes.

Analogous tests applicable to complex-valued random vectors --- rather than time series --- have been descibed by, e.g.,
\cite{SchreierScharfHanssen06}
and %
\cite{WaldenRD09}. However, we need to consider new methodology suitable for very limited degrees of freedom.  
Our test uses the associated spectral matrix which can be estimated by multitaper methods using, say, $K$ tapers. Standard asymptotic distributions for the test statistic are of no use since they would require $K \rightarrow \infty,$
but,
as  $K$ increases so does resolution bandwidth which causes spectral blurring. In many analyses $K$ is necessarily  kept small, and hence our efforts are directed at practical and accurate methodology for hypothesis testing for small $K.$  Our generalized likelihood ratio statistic combined with exact cumulant matching 
gives very accurate rejection percentages and outperforms competitor methods. 

For the scalar case, $(p=1),$ a parametric hypothesis test for propriety of complex time series is given in \cite{Sykulski_etal15a}, \cite{Sykulski_etal15b}.   This is based on the series being well-modelled by a 
Mat{\'e}rn process in  \cite{Sykulski_etal15a} or complex autoregressive process of order one in \cite{Sykulski_etal15b}, and utilises the $\chi^2$ distribution for the test statistic, an asymptotic result.  This is in contrast to our approach which (i)  is suitable for $p>1, $ (ii) is nonparametric, so does not rely on a good fit to a parametric model, and (iii) develops a suitable non-asymptotic distribution for the test statistic. 

Our test  
statistic  is comprised of canonical coherencies arising from the complex-valued vector time series, analogous to the situation for complex-valued random vectors. Canonical analysis of real-valued vector time series has been extensively studied and utilised 
(e.g., \cite{MinTsay05, Reinsel97}), mostly in the context of parametric autoregressive moving-average (ARMA) models. Miyata \cite{Miyata70} looked at real-valued vector time series, and developed canonical correlations through linear functions of discrete vector Fourier transforms of two sets of time series. Rather than work with the Fourier transforms, which are sample values, we instead work with the orthogonal processes underlying the complex-valued vector time series, and whose variances and cross-covariances correspond exactly to the spectral components. We are thus able to define population --- as well as sample --- canonical coherencies for complex-valued vector time series.

Our methodology is demonstrated on ocean current data collected at different depths in the Labrador Sea.

\subsection{Contributions}

Following some background in Section~\ref{sec:backg} on complex-valued time series, and the statistical properties of their spectral matrix estimators  under the Gaussian stationary assumption for $\{ \bZ_t\},$ the contributions of this paper are as follows: 
\begin{enumerate}
\item
In Section~\ref{sec:cancoh} we formally derive the canonical coherencies for $\{ \bZ_t \}$ and $\{\bZ_t^*\}$  and show in Section~\ref{sec:TestPropriety} how a test statistic $T(f)$ for testing ${\bR}_{\bZ}(f)={\bf 0}$ arises from the sample canonical coherencies.
\item
After giving further research context in Section~\ref{sec:context}, 
we carefully study the statistical properties of $M(f)=-2K \log T(f)$ in Section~\ref{sec:BasicP}, concentrating on the small $K$ case. We show that Box's scaled chi-square approximation is exact for $p=1$ but not for $p>1,$
and we derive the cumulants of $M(f).$  
\item
In Section~\ref{sec:Other} we show that for $p>1$ and small $K$ matching the first three cumulants of $M(f)$ exactly to a scaled $F$ distribution performs at least as well as competitor methods.
\item
A simulation study is given in Section~\ref{sec:SimulationT} which supports the use of the scaled $F$ approximation for $M(f)$ for the complex-valued vector time series setting. A data analysis using 6-vector valued oceanographic time series is given in Section~\ref{sec:realdata} which shows that when propriety is rejected, the frequency domain approach usefully shows which frequency bands cause the rejection, which may be linked to the physical processes involved.
\item
In Section~\ref{sec:othervc} we show how our use of canonical coherencies in the complex-valued setting is quite different to an existing approach in the literature derived for real-valued processes, even though there are some structural features in common. 
\end{enumerate}

\section{Background}\label{sec:backg}
\subsection{Some Definitions}
We consider a complex-valued {\it $p$-vector-valued
discrete time stochastic process\/} $\{ {\bZ}_t \}$ 
whose $t$th element, $t \in \mathbb{Z},$ is the column vector 
${\bZ}_t=[Z_{1,t},\ldots,Z_{p,t}]^T,$ 
and  without loss of generality take each component process to have zero mean. The sample interval is $\deltt$ and the Nyquist frequency is $\fN=1/(2\deltt).$
We assume the $p$ processes are {\it jointly} second-order stationary (SOS), i.e., $\cov\{ Z_{l,t+\tau},
Z_{m,t} \}=E\{ Z_{l,t+\tau} Z^*_{m,t}\}$ and
$\rel\{ Z_{l,t+\tau}, Z_{m,t} \}=E\{ Z_{l,t+\tau} Z_{m,t}\},$ $1\leq l,m \leq p,$ are functions of
$\tau$ only. Note that $\rel\{ Z_{l,t+\tau}, Z_{m,t} \}=\cov\{ Z_{l,t+\tau},
Z_{m,t}^*\},$ the covariance between one process and the complex conjugate of the other.

A matrix autocovariance sequence  is then given by 
$
\bs_{\bZ,\tau} 
=E\{ {\bZ}_{t+\tau} {\bZ}_{t}^H \},\,
\tau \in \mathbb{Z},
$
where superscript $H$ denotes Hermitian (complex-conjugate) transpose. 
We define $s_{\bZ,lm,\tau}\equiv (\bs_{\bZ,\tau})_{lm},$
and a matrix
cross-relation sequence follows as
$
\br_{\bZ,\tau}=E\{ {\bZ}_{t+\tau} {\bZ}_{t}^T \},\,
\tau \in \mathbb{Z},
$ 
with $r_{\bZ,lm,\tau}\equiv(\br_{\bZ,\tau})_{lm}.$ From their definitions we see that
$$
s_{\bZ,lm,\tau}=s_{\bZ,ml,-\tau}^*;\,\, r_{\bZ,lm,\tau}=r_{\bZ,ml,-\tau},\,1 \leq l,m \leq p.
$$
We assume 
$
\sum_{\tau=-\infty}^\infty |s_{\bZ,lm,\tau}| < \infty
$ and 
$
\sum_{\tau=-\infty}^\infty |r_{\bZ,lm,\tau}| < \infty,
$
for $1 \leq l \leq m \leq p,$
which means that the Fourier transforms
$S_{\bZ,lm}(f)$ and $R_{\bZ,lm}(f)$ for $1 \leq l,m \leq p,$ exist and are bounded and continuous. In fact for $|f|\leq \fN,$ the corresponding matrices are defined as
\begin{eqnarray*}
\bS_\bZ(f) &=&\deltt \sum_{\tau=-\infty}^\infty \bs_{\bZ,\tau}\,{\e}^{-{\eye} 2\pi f\tau \,\deltt} \quad \mbox{and}\\
\bR_\bZ(f) &=& \deltt\sum_{\tau=-\infty}^\infty
\br_{\bZ,\tau} \e^{-{\eye}2\pi f \tau \deltt}.
\end{eqnarray*}
We note that 
\begin{eqnarray}\label{eq:Rposneg}
\br_{\bZ,\tau}=\br^T_{\bZ,-\tau}&\Longrightarrow& \bR_\bZ(f) =\bR_\bZ^T(-f),
\end{eqnarray}
a result which will prove useful later.

The covariance stationarity means that there exists \cite[p.~317]{Yaglom87} an orthogonal process $\bZ(f)$ such that
$$
\bZ_t=\int_{-1/2}^{1/2} \e^{\eye 2 \pi ft} \dif \bZ(f)
$$ 
where 
$$
E\{ \bZ(f')\bZ^H(f) \}=
\begin{cases}
\bS_\bZ(f) \dif f, &f=f'\\
0, &\text{otherwise}.
\end{cases}
$$

\subsection{Proper Processes}\label{subsec:defproper}
If $\br_{\bZ,\tau}={\bf 0}\,\,\text{for all}\, \tau\in {\mathbb Z},$ or  ${\bR}_{\bZ}(f)={\bf 0}\,\,\text{for all}\, |f|\leq \fN,$ then the process $\{ \bZ_t\}$ is said to be {\it proper}. Equivalently we see that if $\{\bZ_t\}$ is uncorrelated with its complex conjugate $\{ \bZ^*_t\},$ then the vector-valued process is proper. 
This paper considers the problem of  testing that the vector process is proper. 
\begin{remark}
Based on the naming convention adopted in \cite[p.~41]{SchreierScharf10} for complex-valued vectors, an alternative would be to call the component processes `jointly proper.'
\end{remark}

\subsection{Spectral Matrices}

Let \begin{equation}\label{eq:constructZ}
Z_{l,t}= X_{l,t} + \ri Y_{l,t},
\end{equation} 
with $\{ X_{l,t} \}$ and $\{ Y_{l,t} \}$ real-valued, for $l=1,\ldots,p,$ where
$\bv_{t}=[\bX_{t}^{T},\bY^{T}_{t}]^{T}=[X_{1,t},\ldots,X_{p,t},Y_{1,t},\ldots,Y_{p,t}]^{T}$ is a real $2p$-dimensional vector-valued Gaussian stationary process.
Then if 
\begin{equation}\label{eq:defT}
{\bT}\EqualDef 
\left[
\begin{matrix}
{\bI}_p & \eye {\bI}_p \\
{\bI}_p & -\eye {\bI}_p
\end{matrix}
\right],
\end{equation}
we see that
\begin{equation}\label{eq:Zcheck}
{\bT} \bv_t
=
\left[
\begin{matrix}
 {\bX}_t+ \eye {\bY}_t \\ 
{\bX}_t - \eye {\bY}_t 
\end{matrix}\right]
=
\left[
\begin{matrix}
 {\bZ}_t\\ 
{\bZ}_t^* 
\end{matrix}\right]=\bu_t,
\end{equation}
where $\bu_{t}=[\bZ_{t}^{T},\bZ^{H}_{t}]^{T}=[Z_{1,t},\ldots,Z_{p,t},Z_{1,t}^*,\ldots,Z_{p,t}^*]^{T}$ is a real $2p$-dimensional vector-valued Gaussian stationary process.

The spectral matrix for $\bv_t$ is given by 
\begin{equation}\label{eqn:svf}
{\bS}_{\mbv}(f)=\begin{bmatrix}\bS_{\bX\bX}(f) & \bS_{\bX\bY}(f) \\
\bS_{\bY\bX}(f) & \bS_{\bY\bY}(f) \\
\end{bmatrix}\in {\mathbb{C}}^{2p \times 2p}.
\end{equation}
The spectral matrix for $\bu_t$ is $\bS_{\mbu}(f)=\bT\bS_{\mbv}(f)\bT^{H}$ and has the form
\begin{equation}\label{eqn:suf}
{\bS}_{\mbu}(f)=\begin{bmatrix}\bS_{\mbZ}(f) & \bR_{\mbZ}(f) \\
\bR^{H}_{\mbZ}(f) & \bS^{T}_{\mbZ}(-f) \\
\end{bmatrix}\in {\mathbb{C}}^{2p \times 2p}.
\end{equation}
The matrix ${\bS}_{\mbu}(f)$ can be written in the alternative covariance matrix form
\begin{equation*}
E\{ {\boldsymbol U}(f)\, {\boldsymbol U}^H(f)\}={\bS}_{\mbu}(f) \dif f,
\end{equation*}
where 
\begin{equation}\label{eq:defdU}
{\boldsymbol U}(f)\EqualDef [ \dif\mbZ^T(f), \dif\mbZ^H(-f)]^T.
\end{equation}

\subsection{Estimation}
Given a length-$N$ sample $\bv_{0},\ldots,\bv_{N-1}$, form $h_{k,t}\bv_{t}$ using a suitable set of $K$ length-$N$ orthonormal data taper sequences $\{h_{k,t}\},k=0,\ldots,K-1$, and compute 
\begin{equation*}
\bJ_{\mbv,k}(f)=\deltt^{1/2}\sum_{t=0}^{N-1}h_{k,t}\bv_{t}{\rm e}^{{-\rm i} 2 \pi f t \deltt}.
\end{equation*}
In this work we use sine tapers (e.g., \cite{Waldenetal95}).

As $N \rightarrow \infty$, with the number of degrees of freedom, $K$ fixed, 
and with the given taper properties, $\{\bJ_{\mbv,k}(f), k=0,1,\ldots,K-1\}$ are proper, independent and identically distributed random vectors such that
\begin{equation}\label{eqn:jvkfind}
\bJ_{\mbv,k}(f)\EqualDist {\cal N}^{C}_{2p}({\mathbf{0}},\bS_{\mbv}(f)),  \hspace{2mm}0<|f|<{\fN},
\end{equation}
for $k=0,\ldots,K-1$ (e.g., \cite{ChandnaWalden11}). 
As $\bJ_{\mbu,k}(f)=\bT\bJ_{\mbv,k}(f)$, as $N \rightarrow \infty$, with $K$ fixed, $\{\bJ_{\mbu,k}(f), k=0,1,\ldots,K-1\}$ are also a set of proper, independent and identically distributed random vectors each of which are distributed as
\begin{equation}
\bJ_{\mbu,k}(f)\EqualDef {\cal N}^{C}_{2p}({\mathbf{0}},\bS_{\mbu}(f)),  \hspace{2mm}0<|f|<\fN.
\end{equation}

The probability density function (PDF) of $\bJ_{\mbu,k}(f)$ --- a  proper Gaussian vector in ${\mathbb{C}}^{2p}$ is given by \cite{Picinbono96}
\begin{equation}\label{eqn:complexnorm2p}
{\pi^{-p}}[\det\{{\bS}_{\mbu}(f)\}]^{-1}\exp\left\{-{\bJ}_{\mbu,k}^{H}(f){\bS}^{-1}_{\mbu}(f){\bJ}_{\mbu,k}(f)\right\}.
\end{equation}

The independence of $\bJ_{\mbu,k}(f)$'s allows us to write the joint PDF of $\bJ_{\mbu,0}(f), \ldots,\bJ_{\mbu,K-1}(f)$ as the product of their marginal densities given by (\ref{eqn:complexnorm2p}). So the likelihood function, $g_\bJ(\bS_{\bu}(f)|\bJ_{\mbu,0}(f),\ldots,\bJ_{\mbu,K-1}(f)),$  of $\bS_{\bu}(f)$ given $\bJ_{\mbu,0}(f), \ldots,\bJ_{\mbu,K-1}(f),$ is given by
\begin{equation}\label{eqn:gjorig}
[{\pi^{p}}\det\{{\bS}_{\mbu}(f)\}]^{-K}
\exp\left\{-\sum_{k=0}^{K-1}{\bJ}_{\mbu,k}^{H}(f){\bS}^{-1}_{\mbu}(f){\bJ}_{\mbu,k}(f)\right\}.
\end{equation}
Now $\hat{{\bS}}_{\mbu}(f)$ is the sample covariance matrix of $\{\bJ_{\mbu,k}(f); k=0,1,\ldots,K-1\}$, i.e.,
\begin{equation}\label{eqn:sufest}
\hat{{\bS}}_{\mbu}(f)=\frac{1}{K}\sum_{k=0}^{K-1}{\bJ}_{\mbu,k}(f){\bJ}_{\mbu,k}^{H}(f)=\begin{bmatrix}\hat{\bS}_{\mbZ}(f) & \hat{\bR}_{\mbZ}(f) \\
\hat{\bR}^{H}_{\mbZ}(f) & \hat{\bS}^{T}_{\mbZ}(-f) \\
\end{bmatrix}.
\end{equation}
Noting that the argument of $\exp\{\cdot\}$ in (\ref{eqn:gjorig}) is scalar, and so is equal to its trace, 
and recalling the linearity and cyclicity of the trace operator, we can write
\begin{equation}\label{eqn:gJ}
g_\bJ = [{\pi^{p}}\det\{{\bS}_{\mbu}(f)\}]^{-K}\exp\left\{-{K}\tr\{{\bS}^{-1}_{\mbu}(f)\hat{{\bS}}_{\mbu}(f)\}\right\},
\end{equation}
where dependence of $g$ on its arguments is suppressed for convenience. 

For a finite value of $N$, $\{\bJ_{\mbu,k}(f); k=0,1,\ldots,K-1\}$ are proper random variables with
\begin{equation}
\bJ_{\mbu,k}(f)\EqualDist {\cal N}^{C}_{2p}({\mathbf{0}},\bS_{\mbu}(f)), \quad W_{N}<|f|<\fN-W_{N},
\end{equation}
where $[-W_N,W_N]$ is the extent of the spectral window induced by tapering \cite{ChandnaWalden11}. 
For sine tapers
\begin{equation}\label{eq:WNdef}
W_N= (K+1)/[2(N+1)\deltt],
\end{equation}
(e.g., \cite{Waldenetal95}).
Therefore, in practice, we have to restrict interest to frequencies in the range $W_{N}<|f|<\fN-W_{N}$. 

\section{Canonical Coherencies}\label{sec:cancoh}
The structure of the testing problem is related to measures of coherence between vector-valued processes, 
and so we next turn our attention to the idea of canonical coherence.
\subsection{New Series Defined by Cross-correlations}
Consider the cross-correlation of complex-valued deterministic matrix sequence $\{\bA_t\}$ with the time series $\{ \bZ_t \}$ to give $\{\bxi_t\}:$
$$
\bxi_t=\bA^* \star \bZ_t\EqualDef \sum_{u=-\infty}^\infty
\bA^*_u \bZ_{t+u}.
$$
Likewise we define the cross-correlation of complex-valued deterministic matrix sequence $\{\bB_t\}$ with the time series $\{ \bZ_t^* \}$ to give $\{\beeta_t\}:$
$$
\beeta_t=\bB^* \star \bZ_t^*\EqualDef \sum_{u=-\infty}^\infty
\bB^*_u \bZ_{t+u}^*.
$$
Component-wise we have
\begin{equation}
\begin{bmatrix}
\xi_{1,t} \\
\xi_{2,t} \\
\vdots \\
\xi_{p,t} \\
\end{bmatrix}=\sum_{u}\begin{bmatrix}a^*_{11,u} & \hdots &\hdots  & a^*_{1p,u} \\
a^*_{21,u} & \hdots &\hdots  & a^*_{2p,u} \\
\vdots &  & & \vdots \\
a^*_{p1,u} & \hdots & \hdots & a^*_{pp,u} \\
\end{bmatrix} \begin{bmatrix}Z_{1,t+u} \\
Z_{2,t+u} \\
\vdots \\
Z_{p,t+u} \\
\end{bmatrix}.
\end{equation}
So, for $j=1,\ldots,p$, 
\begin{equation}\label{eqn:zjt}
\xi_{j,t}=\sum_{u}a^*_{j1,u}Z_{1,t+u}+\cdots+\sum_{u}a^*_{jp,u}Z_{p,t+u}.
\end{equation}
The spectral representation theorem allows us to write $\xi_{j,t}, j=1,\ldots,p$ and $Z_{l,t}$, $l=1,\ldots,p,$ as
\begin{equation*}
\xi_{j,t}=\int_{-\fN}^{\fN}\e^{\eye 2\pi ft \,\deltt} \dif Z_{\xi_j}(f); \hspace{2mm} Z_{l,t}=\int_{-\fN}^{\fN} \e^{\eye 2\pi ft \,\deltt}\dif Z_{l}(f) . 
\end{equation*}
Substituting the spectral representation for $Z_{1,t}$ 
in the first term of (\ref{eqn:zjt}), we get
\begin{eqnarray*}
\sum_{u}a^*_{j1,u}Z_{1,t+u}
&=&\int_{-\fN}^{\fN}\e^{\eye 2\pi ft \,\deltt}  A^*_{j1}(f)\,\dif Z_{{1}}(f), 
\end{eqnarray*}
where $A_{jl}(f)=\sum_{u}a_{jl,u}{\rm e}^{-{\eye}2 \pi fu \deltt}$.
Proceeding in analogous fashion, and using the fact that the orthogonal process in a spectral representation is unique \cite[p.~34]{Chonavel02},
we obtain
\begin{eqnarray*}
\dif Z_{\xi_j}(f) &=& A^*_{j1}(f)\dif Z_{1}(f)+\ldots+A^*_{jp}(f)\dif Z_{p}(f)\\
&{\displaystyle{\EqualDef}}& \bA_j^H(f) \dif\bZ(f).
\end{eqnarray*}
So
\begin{equation}
\xi_{j,t} 
=\int_{-\fN}^{\fN}\e^{\eye 2\pi ft \,\deltt}\bA_j^H(f) \dif\bZ(f).\label{eq:firstcan}
\end{equation}
For $\{\beeta_t\}$ a similar procedure gives
\begin{eqnarray*}
\dif Z_{\eta_j}(f) &=& B^*_{j1}(f)\dif Z_{1}^*(-f)+\ldots+B^*_{jp}(f)\dif Z_{p}^*(-f)\\
&{\displaystyle{\EqualDef}}& \bB_j^H(f) \dif\bZ^*(-f),
\end{eqnarray*}
and
\begin{equation}\label{eq:secondcan}
\eta_{j,t}=\int_{-\fN}^{\fN}\e^{\eye 2\pi ft \,\deltt}\bB_j^H(f) \dif\bZ^*(-f).
\end{equation}

The usual definition of the  (magnitude squared) coherencies $\gamma_j^2(f)$ between series $\{\xi_{j,t}\}$ and 
$\{ \eta_{j,t} \}$ is 
\begin{eqnarray*}
\gamma_j^2(f) &=&\frac{|E\{ \dif Z_{\xi_j}(f)\dif Z^H_{\eta_j}(f)\}|^2}{E\{| \dif Z_{\xi_j}(f)|^2\}E\{| \dif Z_{\eta_j}(f)|^2\}}\\
&=&|{\rm corr}\{\dif Z_{\xi_j}(f),\dif Z_{\eta_j}(f)\}|^2.
\end{eqnarray*}

\begin{remark}
It should be emphasized that throughout we use the usual definition of coherence as a magnitude squared quantity, basically a squared correlation coefficient.
\end{remark}

\subsection{Finding Canonical Coherencies}
In vector notation,
\begin{equation}\label{eqn:linrot}
\dif \bZ_{\bxi}(f)\!=\!\bA^H(f) \dif \bZ(f)\,\,\mbox{and}\,\,\dif \bZ_{\beeta}(f)\!=\!\bB^H(f) \dif\bZ^*(-f),
\end{equation}
where $\bA(f)=[\bA_1(f), \bA_2(f),\ldots, \bA_p(f)].$ 

Consider $|{\rm corr}\{\dif Z_{\xi_j}(f),\dif Z_{\eta_j}(f)\}|$. This can be written
\begin{eqnarray*}\label{eq:extra}
&&\!\!\!\!\!\!\!\!\!\!\!\frac{|\bA_j^H(f)\bR_{\mbZ}(f)\dif f\bB_j(f)|}
{[\bA_j^H(f) \bS_{\mbZ}(f)\dif f \bA_j(f)]^{1/2}
[\bB_j^H(f) \bS^T_{\mbZ}(-f)\dif f \bB_j(f)]^{1/2}}\\
&&\!\!\!\!\!\!\!\!\!\!\!= \frac{|\bA_j^H(f)\bR_{\mbZ}(f)\bB_j(f)|}
{[\bA_j^H(f) \bS_{\mbZ}(f) \bA_j(f)]^{1/2}
[\bB_j^H(f) \bS^T_{\mbZ}(-f) \bB_j(f)]^{1/2}}.
\end{eqnarray*}
Suppose we choose $\bA(f)$ and $\bB(f)$ so that 
\begin{equation}\label{eq:standardize}
{\bA^H(f) \bS_{\mbZ}(f) \bA(f)=\bI_p=
\bB^H(f) \bS^T_{\mbZ}(-f) \bB(f)}.
\end{equation}
Then
$$
|{\rm corr}\{\dif Z_{\xi_j}(f),\dif Z_{\eta_j}(f)\}|=
|\bA_j^H(f)\bR_{\mbZ}(f)\bB_j(f)|.
$$

 It also ensures
that for $j\not=k,$
\begin{equation}\label{eq:corra}
{\rm corr}\{\dif Z_{\xi_j}(f),\dif Z_{\xi_k}(f)\}=0=
{\rm corr}\{\dif Z_{\eta_j}(f),\dif Z_{\eta_k}(f)\}.
\end{equation}


Define
$$
\bK(f){\displaystyle\EqualDef}
\bA^H(f)\bR_{\mbZ}(f)\bB(f),
$$
so that 
$$
|K_{jj}(f)|= |{\rm corr}\{\dif Z_{\xi_j}(f),\dif Z_{\eta_j}(f)\}|.
$$

\begin{definition}\label{def:one}
The first definition of the canonical coherence problem under the standardization in (\ref{eq:standardize}) is  as follows. Find $\bA_1(f)$ and $\bB_1(f)$ such that 
$|K_{11}(f)|=|{\rm corr}\{\dif Z_{\xi_1}(f),\dif Z_{\eta_1}(f)\}|$ is maximized. Next find $\bA_2(f)$ and $\bB_2(f)$ such that 
$|K_{22}(f)|=|{\rm corr}\{\dif Z_{\xi_2}(f),\dif Z_{\eta_2}(f)\}|$ is maximized, subject to 
$\dif Z_{\xi_2}(f),\dif Z_{\eta_2}(f)$
being uncorrelated with $\dif Z_{\xi_1}(f),\dif Z_{\eta_1}(f).$
 In general, at step $j$ for $j=2,\ldots,p,$ $\bA_j(f)$ and $\bB_j(f)$ are found such that 
$|K_{jj}(f)|=|{\rm corr}\{\dif Z_{\xi_j}(f),\dif Z_{\eta_j}(f)\}|$ is maximized subject to 
$\dif Z_{\xi_j}(f),\dif Z_{\eta_j}(f)$ being uncorrelated with $\dif Z_{\xi_k}(f),\dif Z_{\eta_k}(f)\}$ for $1\leq k<j.$
\end{definition}

The  problem can be defined in a different but equivalent way  \cite{Schreier08}. 

\begin{definition}
The second definition of the canonical coherence problem under the standardization in (\ref{eq:standardize}) is  as follows.
Choose  $\bA(f)$ and $\bB(f)$ 
such that all partial sums over the $|K_{jj}(f)|$ are maximized, i.e.,
\begin{equation}\label{eq:defineprob}
\max_{\bA(f),\bB(f)} \sum_{j=1}^r 
|K_{jj}(f)|, \, r=1,\ldots,p.
\end{equation}
\end{definition}

\begin{lemma}\label{lem:one}
The canonical coherencies
$$l_j^2(f)\EqualDef|K_{jj}(f))|^2,\quad j=1,\ldots,p$$ 
and $\bA_j(f)$ and $\bB_j(f)$ for $j=1,\ldots,p,$ solving (\ref{eq:defineprob}) are eigenvalues and eigenvectors defined as follows:
\begin{eqnarray*}
\bS_{\mbZ}^{-1}(f)\bR_{\mbZ}(f)\bS^{-T}_{\mbZ}(-f)
\bR_{\mbZ}^H(f)\bA_j(f)\!\!\!&=&\!\!\!l_j^2(f)\bA_j(f)\\
\bS^{-T}_{\mbZ}(-f)\bR_{\mbZ}^H(f)\bS_{\mbZ}^{-1}(f)\bR_{\mbZ}(f)\bB_j(f)\!\!\!&=&\!\!\!l_j^2(f)\bB_j(f).
\end{eqnarray*}
Moreover we have that as a result,
\begin{equation}\label{eq:corrb}
{\rm corr}\{\dif Z_{\xi_j}(f),\dif Z_{\eta_k}(f)\}=0, \,\,\,\mbox{for}\,\,\, j,k=1,\ldots,p; j\not=k.
\end{equation}
\end{lemma}
\begin{IEEEproof}
See Appendix~\ref{app:lem1}.
\end{IEEEproof}
\begin{remark}
From Lemma~\ref{lem:one}
the optimal $\bA_j(f)$ and $\bB_j(f)$  give rise to the $j$th pair of canonical series via (\ref{eq:firstcan}) and (\ref{eq:secondcan}). 
\end{remark}
\begin{remark}
Results (\ref{eq:corra}) and (\ref{eq:corrb}) ensure that the uncorrelated requirements in Definition~\ref{def:one} hold.
\end{remark}

\section{Generalized Likelihood Ratio Test}\label{sec:TestPropriety}

\subsection{Formulation}
The GLR test statistic for
\begin{equation}\label{eq:TSsetting}
H_{0}: \bR_{\mbZ}(f)={\mathbf{0}} \quad\text{versus}\quad H_{1}: \bR_{\mbZ}(f)\neq{\mathbf{0}}, 
\end{equation}
for any $W_{N}<|f|<\fN-W_{N}$, is given by ratio of the likelihood function (\ref{eqn:gJ}) with  ${{\bS}}_{\mbu}(f)$ constrained to have zero off-diagonal blocks ($\bR_{\mbZ}(f)=\mathbf{0}$) to the likelihood function with ${{\bS}}_{\mbu}(f)$ unconstrained, i.e.,
\begin{equation}\label{eqn:GLR}
\frac{\underset{{{\boldsymbol S}}_{\mbu}(f): \boldsymbol R_{\mbZ}(f)=\mathbf{0}}{\max} g_\bJ}{\underset{{{\boldsymbol S}}_{\mbu}(f)}\max \hspace{1mm}g_{\bJ}}\EqualDef L_G(f). 
\end{equation}

The unconstrained maximum likelihood estimate of the covariance matrix ${{\bS}}_{\mbu}(f)$ is given by the corresponding sample covariance matrix $\hat{{\bS}}_{\mbu}(f)$ in (\ref{eqn:sufest}), thus maximum likelihood estimate of ${{\bS}}_{\mbu}(f)$ under the constraint $\bR_{\mbZ}(f)=\mathbf{0}$ is,
\begin{equation}\label{eqn:sufestro}
\breve{{\bS}}_{\mbu}(f)=\begin{bmatrix}\hat{\bS}_{\mbZ}(f) & \mathbf{0} \\
\mathbf{0} & \hat{\bS}^{T}_{\mbZ}(-f) \\
\end{bmatrix}.
\end{equation}
From (\ref{eqn:gJ}), (\ref{eqn:GLR}) it follows that $T(f)\,\, \displaystyle{\EqualDef}\,\, L_G^{1/K}(f)$ is
\begin{eqnarray}\label{eqn:preveqn}
T(f)&=& \frac{[\det\{ \breve{{\bS}}_{\mbu}(f)\}]^{-1}\exp\left\{-\tr\{\breve{{\bS}}^{-1}_{\mbu}(f)\hat{{\bS}}_{\mbu}(f)\}\right\}}
{[\det\{ \hat{{\bS}}_{\mbu}(f)\}]^{-1}\exp\left\{-\tr\{\hat{{\bS}}^{-1}_{\mbu}(f)\hat{{\bS}}_{\mbu}(f)\}\right\}} \nonumber \\ 
&=&\det\{\breve{{\bS}}^{-1}_{\mbu}(f)\hat{{\bS}}_{\mbu}(f)\}\nonumber\\
&\times&\exp\left\{-\tr\{\breve{{\bS}}^{-1}_{\mbu}(f)\hat{{\bS}}_{\mbu}(f)-\bI_{2p}\}\right\}.
\end{eqnarray}

The result (\ref{eqn:gJ}) is valid for $W_{N}<|f|<\fN-W_{N},$ but
from (\ref{eq:Rposneg}) we see that if $\bR_{\mbZ}(f)={\mathbf{0}}$ for $f>0$ then it is also ${\mathbf{0}}$
for $f<0.$ Hence in practice we need only concern ourselves with the positive frequency range
$W_{N}< f<\fN-W_{N},$ and calculate $T(f)$ over this interval.

 From (\ref{eqn:sufest}) and (\ref{eqn:sufestro}) we see that
\begin{equation*}
\breve{{\bS}}^{-1}_{\mbu}(f)\hat{{\bS}}_{\mbu}(f)=\begin{bmatrix}\bI_{p} & \hat{\bS}^{-1}_{\mbZ}(f)\hat{\bR}_{\mbZ}(f) \\
\hat{\bS}^{-T}_{\mbZ}(f)\hat{\bR}^{H}_{\mbZ}(f) & \bI_{p} 
\end{bmatrix},
\end{equation*}
so that the $\exp\{\cdot\}$ term is unity.
\allowdisplaybreaks
Thus (\ref{eqn:preveqn}) becomes
\begin{eqnarray}
\!\!\!\!\!\!\!\!\!\!T(f)&=&\det\{\breve{{\bS}}^{-1}_{\mbu}(f)\hat{{\bS}}_{\mbu}(f)\}\label{eqn:allU}\\
&=&
\det\left
\{\begin{bmatrix}\bI_{p} & \hat{\bS}^{-1}_{\mbZ}(f)\hat{\bR}_{\mbZ}(f) \\
\hat{\bS}^{-T}_{\mbZ}(-f)\hat{\bR}^{H}_{\mbZ}(f) & \bI_{p} 
\end{bmatrix}\right\}\nonumber\\
&=& \det\{\bI_{p}-\hat{\bS}^{-1}_{\mbZ}(f)\hat{\bR}_{\mbZ}(f)\hat{\bS}^{-T}_{\mbZ}(-f)\hat{\bR}^{H}_{\mbZ}(f) \}\label{eqn:glrform}\\
&=&\frac{ \det\{\hat{\bS}_{\mbZ}(f)-\hat{\bR}_{\mbZ}(f)\hat{\bS}^{-T}_{\mbZ}(-f)\hat{\bR}^{H}_{\mbZ}(f) \}}{\det \{\hat{\bS}_{\mbZ}(f)\}}.\nonumber
\end{eqnarray}
Starting with (\ref{eqn:allU}) and using
(\ref{eqn:sufestro}) we also have that
\begin{equation}\label{eqn:glrformtwo}
T(f)=\frac{\det\{\hat{{\bS}}_{\mbu}(f)\}}{\det\{ \breve{{\bS}}_{\mbu}(f)\}}
=
\frac{\det\{\hat{{\bS}}_{\mbu}(f)\}}{\det\{
\hat{\bS}_{\mbZ}(f) \}\det\{ \hat{\bS}_{\mbZ}(-f)
 \}}.
\end{equation}
Now, the GLR test may be based on any of the above equivalent forms for $T(f).$
Form (\ref{eqn:glrformtwo}), unlike other formulations  does not involve computation of either $\hat{\bS}^{-1}_{\mbZ}(f)$ or $\hat{\bS}^{-T}_{\mbZ}(-f)$.

By definition of the GLR test statistic (\ref{eqn:GLR}), we shall reject the null hypothesis of $\bR_{\bZ}(f)=\mathbf{0}$, for small values of $T(f).$
For a given size $\alpha$, the rule is to reject $\bH_{0}$ iff
\begin{equation}
T(f;N,K,p)\leq c,
\end{equation}
where Pr$(T(f;N,K,p)\leq c| H_{0})=\alpha$.
Here we have used the more precise notation $T(f;N,K,p)$ which emphasizes the dependence of the GLR test on (i) the sample size $N$, (ii) the number of tapers $K$ (also the number of complex degrees of freedom), and (iii) dimension $p$ of the complex time series.

\subsection{Invariance}
Now 
$$\bR_{\mbZ}(f) \dif f\EqualDef E\{\dif \bZ(f)\dif\bZ^{T}(-f)\}.
$$
Apply ${\bL}(f)\in {\mathbb{C}}^{p \times p}$
to $\dif \bZ(f)$ so that $\dif \bZ(f) \rightarrow {\bL}(f)\dif \bZ(f),$
and therefore $\dif\bZ^{T}(-f) \rightarrow 
{\bL}^*(-f)\dif\bZ^{T}(-f).$ So 
\begin{eqnarray*}
\bR_{\mbZ}(f) &=&{\mathbf{0} }\implies
E\{ {\bL}(f)\dif \bZ(f) [{\bL}^*(-f)\dif\bZ^{T}(-f)]^H\}\\
&=&{\bL}(f) \bR_{\mbZ}(f)\dif f {\bL}^T(-f)
={\mathbf{0} },
\end{eqnarray*}
i.e., $\bR_{\mbZ}(f)=\mathbf{0}$ is invariant to the linear transformation $\dif \bZ(f) \rightarrow {\bL}(f)\dif \bZ(f)$. So  the decision rule for our GLR test must be likewise invariant.

Note that under this transformation,
\begin{eqnarray*}
\mbu(f)&\rightarrow& [ {\bL}(f)\dif\mbZ(f), {\bL}^*(-f)\dif\mbZ^*(-f)]^T\\
&=& \left[
\begin{matrix}
\bL(f) & {\bf 0}\\
{\bf 0} & \bL^*(-f)
\end{matrix}\right]
\mbu(f)\\
&{\displaystyle \EqualDef} & \bQ(f) \mbu(f),
\end{eqnarray*}
so that we require invariance under the group action 
${\bS}_{\mbu}(f)\rightarrow {\bQ}(f){\bS}_{\mbu}(f){\bQ}^{H}(f)$.

Under the null hypothesis ${\bS}_{\mbu}(f)$ takes the form in (\ref{eqn:sufestro}) so that
${\bQ}(f){\bS}_{\mbu}(f){\bQ}^{H}(f)$ is
$$
\left[
\begin{matrix}
\bL(f) \bS_{\mbZ}(f) \bL^H(f) & {\bf 0}\\
{\bf 0} & \bL^*(-f) \bS^*_{\mbZ}(-f) \bL^T(-f)
\end{matrix}
\right],
$$
and the choice $\bL(f)=\bS_{\mbZ}^{-1/2}(f)$ 
(which exists for $\bS_{\mbZ}(f)$ positive definite) renders the matrix equal  to ${\bI}_{2p}.$ This means that under the null hypothesis we can always replace ${\bS}_{\mbu}(f)$ by ${\bI}_{2p}$ without loss of generality.

From Lemma~\ref{lem:one}  we know that the eigenvalues $l_j^2(f)$ of ${{\bS}}^{-1}_{\mbZ}(f){{\bR}}_{\mbZ}(f){{\bS}}^{-T}_{\mbZ}(-f){{\bR}}^{H}_{\mbZ}(f)$ are canonical coherencies which are invariant under the group action specified above; moreover, the corresponding empirical or sample canonical coherencies are maximal invariant and the GLR statistic --- which requires this invariance --- must be a function of them.    

Let ${\ell}_{j}^2(f), j=1,\ldots,p,$ be the sample versions of the 
canonical coherencies $l_j^2(f)$ between
$\dif\mbZ(f)$ and $ \dif\mbZ^*(-f).$
They are the sample eigenvalues of $\hat{{\bS}}^{-1}_{\mbZ}(f)\hat{{\bR}}_{\mbZ}(f)\hat{{\bS}}^{-T}_{\mbZ}(-f)\hat{{\bR}}^{H}_{\mbZ}(f)$. Then from (\ref{eqn:glrform}) it follows that for $W_{N}< f<\fN-W_{N}$,
\begin{eqnarray}
\!\!\!\!\!\!\!\!\!\!\!\!\!T(f)&=& \det(\bI_{p}-\hat{\bS}^{-1}_{\mbZ}(f)\hat{\bR}_{\mbZ}(f)\hat{\bS}^{-T}_{\mbZ}(-f)\hat{\bR}^{H}_{\mbZ}(f)) \label{eqn:firstone}\\   &=&\prod_{j=1}^{p}(1-{\ell}^{2}_{j}(f)). \label{eqn:lrtfreq}
\end{eqnarray}

\section{Research Context}\label{sec:context}
Testing $\bR_{\bZ}(f)=\mathbf{0}$ is the same as  testing the independence of two complex Gaussian $p$-vectors, namely 
$\dif\mbZ(f)$ and $ \dif\mbZ^*(-f),$ (see (\ref{eq:defdU})).
The GLR test based on (\ref{eqn:glrformtwo}) falls in the class of multiple independence tests in multivariate statistics theory. Some distributional results for the complex case were given in \cite{Krishnaiah_etal76} but did not include the case of interest here, namely two  $p$-vectors.  A later paper \cite{Fang_etal82} gave the exact distribution of a power of $T(f)$ but this involves an infinite sum with very complicated components; small $K$ approximations were not discussed. Other relevant results can be found  in \cite{Gupta71} and \cite{Krishnaiah_etal83}, and these are discussed in detail in Section~\ref{subsec:beta}.

The statistic $T(f)$ is the frequency-domain time series analogue to those used in \cite{OllilaKoivunen04,SchreierScharfHanssen06} and
\cite{WaldenRD09} to examine independence between a Gaussian random vector and its complex conjugate. In \cite{OllilaKoivunen04,SchreierScharfHanssen06} a complex formulation was maintained but only an asymptotic approach to testing was considered. In \cite{WaldenRD09} a real-valued representation of the problem was used and Box's scaled chi-square method was used to improve on the asymptotic critical values. In the rest of this paper we adopt the complex formulation, derive Box's refinement, but also improve on it for $p>1$ by {\it exactly} matching the first three cumulants  to a scaled $F$-distribution. (We point out that Box's refinement is exact for $p=1.$) This latter $F$-method is very simple to implement practically, involving only the first three polygamma functions.

We emphasize that our efforts are directed at practical and accurate methodology for small $K.$ This is important in a time series setting where as  $K$ increases so does resolution bandwidth which potentially causes spectral blurring. In many analyses $K$ must necessarily be kept small.
In the remainder of this paper  we will always assume any frequency under consideration to lie in the interval 
$W_{N}< f<f_{N}-W_{N}.$

\section{Basic  Properties of Test Statistic}\label{sec:BasicP}
\subsection{Asymptotic Behaviour}
The application of Wilk's theorem \cite[p.~132]{YoungSmith05} gives that under $H_{0},$ as $K\rightarrow \infty$, 
\begin{equation}\label{eqn:Wilks}
M(f)\EqualDef-2\log L_G(f)=-2K\log T(f) \overset{{\rm d}}\rightarrow \chi^2_{\nu}
\end{equation}
where $\overset{\rm d}\rightarrow $ denotes convergence in distribution and $\chi^2_{\nu}$ denotes the chi-square distribution with $\nu$ degrees of freedom. Here $\nu$ is the difference between the number of free real parameters under $H_{0}$ and $H_{1}$. Comparing $\breve{{\bS}}_{\mbu}(f)$ in
(\ref{eqn:sufestro}) (for $H_0$) and
${\bS}_{\mbu}(f)$ in (\ref{eqn:suf}) (for $H_1$) we note that $\bR^{H}_{\mbZ}(f)$ follows directly from
$\bR_{\mbZ}(f)$ so that there is only an additional $2p^2$ degrees of freedom, i.e., those contributed by $\bR_{\mbZ}(f).$ Hence we have $\nu=2p^2.$

While (\ref{eqn:Wilks}) is a very useful and convenient result when the exact distribution of the GLR test statistic is analytically intractable,
$K$ here denotes the number of tapers used for multitaper spectral estimation and not the sample size $N$.
For a given value of $N$, $K$ could be around $10$ or less. Since (\ref{eqn:Wilks}) is an asymptotic result, $K$ must be sufficiently large to expect a reasonable $\chi^2_{\nu}$ approximation to $-2K\log T(f).$ Since $K$ may not be large in a time series setting, a small-$K$ approximation to the distribution of the test statistic under the null hypothesis is imperative.

\subsection{Moments}\label{sec:Box}
Since $\bJ_{\bu,k}(f), k=0,\ldots,K-1,$ are Gaussian distributed random vectors, from (\ref{eqn:sufest}) it follows that  
\begin{equation}\label{eqn:ACWishart}
\bA \EqualDef K\hat{\bS}_{\bu}(f) \EqualDist {\cal W}^{C}_{2p}(K, {\bS}_{\bu}(f)),
\end{equation}
i.e., $\bA(f)$ is distributed as a $2p$-dimensional complex Wishart distribution with $K$ complex degrees of freedom and mean $K {\bS}_{\bu}(f).$
Given the form of $\hat{\bS}_{\bu}(f)$, we partition $\bA(f)$ analogously in terms of sub-matrices as
\begin{equation}
\bA(f)=\begin{bmatrix} \bA_{11}(f) & \bA_{12}(f)\\
\bA_{21}(f) & \bA_{22}(f) \end{bmatrix}. 
\end{equation}
Then the GLR test statistic in (\ref{eqn:glrformtwo}) can be expressed as
\begin{equation}\label{eqn:formtouse}
L_G^{1/K}(f)=\frac{\det\{\bA(f)\}}{\det\{\bA_{11}(f)\}\det\{\bA_{22}(f)\}}.
\end{equation}

\begin{lemma}\label{lemma:one}
The $r$th moment of $L_G(f),$ namely $E\{ L_G^r(f) \},$ is given by
\begin{equation}\label{eqn:moment}
\frac{\prod_{j=1}^{p}\Gamma(K-j+1)}{\prod_{j=1}^{p}\Gamma(K-j-p+1)}
\frac{\prod_{j=1}^{p}\Gamma(K[1+r]-j-p+1)}{\prod_{j=1}^{p}\Gamma(K[1+r]-j+1)}.
\end{equation}
\end{lemma}
\begin{IEEEproof}
This is given in Appendix~\ref{app:lemmaone}.
\end{IEEEproof}
A random variable $0\leq W\leq 1$ is said to be of Box-type \cite[eqn.~(70)]{Box49} if for all
$r\in {\mathbb N},$
\begin{equation}\label{eqn:boxform}
E\{ W^r \}=C_{0}\left[\frac{\prod_{j=1}^{l}b^{b_{j}}_{j}}{\prod_{i=1}^{m}a^{a_{i}}_{i}}\right]^{r}\frac{\prod_{i=1}^{m}\Gamma(a_{i}[1+r]+\vartheta_{i})}
{\prod_{j=1}^{l}\Gamma(b_{j}[1+r]+\zeta_{j})},
\end{equation}
where $\sum_{i=1}^{m}a_{i}=\sum_{j=1}^{l}b_{j}$, and the constant term $C_{0}$ is 
$$
C_0=\frac
{ \prod_{j=1}^{l}\Gamma(b_{j}+\zeta_{j}) }
{ \prod_{i=1}^{m}\Gamma(a_{i}+\vartheta_{i}) },
$$
so that it's zero'th moment is unity.

We see that $L_G(f)$ is a random variable of Box-type with 
$$
m=l=p;\, a_{i}=K;\, b_{j}=K;\, \vartheta_{i}=1-i-p,\, \zeta_{j}=1-j,
$$
and $C_0$ is 
$$
C_0=\prod_{j=1}^p \frac{\Gamma(K-j+1)}{\Gamma(K-j-p+1)}.
$$

\subsection{Cumulants}
The moment generating function for $M(f)=-2\log L_G(f)$ is given by (with $f$ suppressed),
$\phi_M(s)=E\{ \re^{sM}\}=E\{ L_G^{-2s}\}$
so using (\ref{eqn:moment}),
$$
\phi_M(s)=C_0\prod_{j=1}^{p}\frac{\Gamma(K[1-2s]-j-p+1)}{\Gamma(K[1-2s]-j+1)}.
$$
The Gamma functions will be
valid if $-2Ks+K-j-p+1>0$ for all $j=1,\ldots,p,$
which requires $-2s>(2p-1-K)/K.$

The cumulants $\kappa_i$ of $M$ can be easily obtained from the cumulant generating function by successively differentiating $\log\phi_M(s)$
and setting $s=0.$ Notice that the requirement $-2s>(2p-1-K)/K$ corresponds to
$K\geq 2p$ when $s=0.$ Then, for $i\geq 1,$
$$
\kappa_i= \left. \frac{\dif^i \log\phi_M(s)}{(\dif s)^i}
\right|_{s=0} 
$$
so that $\kappa_i$ is
\begin{equation}\label{eq:cums}
[-2K]^i\sum_{j=1}^{p}
\!\!\left[\psi^{(i-1)}(K-j-p+1)-\psi^{(i-1)}(K-j+1)\right].
\end{equation}
Here for $i=1$, $\psi(x)=[\dif \log\Gamma(x)]/{\dif x}$ is the digamma function, while for $i=2$ and 3, $\psi^{(1)}(x)$ and $\psi^{(2)}(x)$ are the trigamma and tetragamma functions respectively; these are all `polygamma 
functions.' 
$\kappa_1$ is the mean, $\kappa_2$ is the variance, $\kappa_3/\kappa_2^{3/2}$ is the skewness and
$\kappa_4/\kappa_2^{2}$ is the excess kurtosis. 

\subsection{Scaled chi-square approximation}\label{subsec:Boxapprox}

Box \cite{Box49} provides a scaled chi-squared approximation for $M$ 
of the form
$M(f) \,{\displaystyle{\EqualDist}}\, c_B \chi^2_{d}.$ The constant $c_B$ is chosen so that the cumulants of $c_B \chi^2_{d}$ match those of $M(f)$ up to an error of order $O(K^{-2}).$
The degrees of freedom $d$ associated with the chi-square approximation for $M(f)$ is given by Box \cite{Box49}
\begin{eqnarray*}
d&=&-2\left[\sum_{i=1}^{p}\vartheta_{i}-\sum_{j=1}^{p}\zeta_{j}\right]\\
&=&-2\left[\sum_{i=1}^{p}(1-i-p)-\sum_{j=1}^{p}(1-j)\right]\\
&=&-2\left[-\sum_{i=1}^{p}i-\sum_{i=1}^{p}p+\sum_{j=1}^{p}j\right]=2p^{2}=\nu,
\end{eqnarray*}
as expected.
The scaling factor $c_{B}$ is a constant determined as follows \cite[p.~338]{Box49}. Define
\begin{equation}\label{eq:Bern}
\omega_{n}=\frac{(-1)^{n+1}}{n(n+1)} \left[\sum_{i=1}^{p}\frac{B_{n+1}(\vartheta_{i})}{a^{n}_{i}} -\sum_{j=1}^{p}\frac{B_{n+1}(\zeta_{j})}{b^{n}_{j}}\right]
\end{equation}
where $B_{n}(x)$ is the Bernoulli polynomial of degree $n$ and order unity, with
\begin{equation*}
B_{2}(x)=x^2-x+\frac{1}{6};\quad B_{3}(x)=x^3-\frac{3}{2}x^2+\frac{1}{2}x.
\end{equation*}
Subsequently, let $W_{1}=2\omega_{1}/d$ and $W_{2}=4\omega_{2}/d$, then $c_{B}$ is chosen according to the following rule:
\begin{equation*}
c_{B}=
\begin{cases}
(1-W_{1})^{-1} & \mbox{if}\,\, W_{2}\geq W^{2}_{1}\\
1+W_{1} & \mbox{otherwise}.
\end{cases}
\end{equation*}
Using (\ref{eq:Bern}) we find that
\begin{equation*}
W_{1}=\frac{p}{K};\quad 
W_{2}=\frac{(7{p^2}-1)}{6 K^2}.
\end{equation*}
It is straightforward to see that $W_{2}\geq W^{2}_{1}$ for all $(K,p)$ combinations, implying that 
$c_{B}=K/(K-p)$, giving Box's finite sample approximation as
\begin{equation}\label{eqn:BoxApprox}
M(f)\EqualDist \frac{K}{K-p}\chi_{2p^{2}}^{2}.
\end{equation}
(This agrees with (\ref{eqn:Wilks}) asymptotically as $K\rightarrow \infty$ for a fixed dimension $p.$)

For the case $p=1$ we note that $T(f)$ in (\ref{eqn:firstone}) becomes
$$
T(f)=1-\frac{| {\hat R}_Z(f)|^2}{{\hat S}_Z(f){\hat S}_Z(-f)}
=1-{\hat\gamma^2}_*(f)
$$
where ${\hat\gamma^2}_*(f)$ is the `conjugate coherence,'
i.e., the ordinary coherence between $\{Z_t\}$ and $\{ Z_t^*\}$ (e.g., 
\cite{ChandnaWalden11}). 
Then $M(f)=-2K\log(1-{\hat\gamma^2}_*(f)).$ Under the null hypothesis it is known that 
\begin{equation}\label{eq:cohdist}
{\hat\gamma^2}_*(f) \EqualDist  {\rm beta}(1,K-1),
\end{equation}
i.e., coherence has the ${\rm beta}(1,K-1)$ distribution. It then follows readily that $M(f)$ has PDF
$$
f_M(x)=\frac{K-1}{2K} \re^{-x \left[\frac{K-1}{2K}\right]},
$$
so that $M(f) \,{\displaystyle{\EqualDist}}\, \frac{K}{K-1} \chi^2_{2}$
and Box's approximation (\ref{eqn:BoxApprox}) is in fact {\it exact}\/ for the case $p=1.$ When $p=1$ we note that $W_2=W_1^2.$
\begin{remark}
For small values of $K$, 
matching cumulants  of $M(f)$ up to an error of order $O(K^{-2})$
 could be problematic for $p>1$ \cite[p.~329]{Box49}.
This leads us to consider other approaches. 
\end{remark}

\section{Other Statistical Approaches}\label{sec:Other}

\subsection{Product of Independent Beta Random Variables}\label{subsec:beta}
\begin{lemma}\label{lemma:two}
Under the null hypothesis
the distribution of $T(f)$ can be expressed as a product of independent beta random variables: 
\begin{equation}\label{eq:prodbeta}
T(f)\EqualDist \prod_{j=1}^p B_j,
\end{equation}
where $B_j {\displaystyle\EqualDist}\,\, {\rm beta}(K+1-j-p,p),$ independently.
\end{lemma}
\begin{IEEEproof}
This is given in Appendix~\ref{app:lemmatwo}.
\end{IEEEproof}
\begin{remark}
If $p=1,$ (\ref{eq:prodbeta}) gives $T(f)\,\,{\displaystyle\EqualDist}\,\,{\rm beta}(K-1,1),$ as it should since $T(f)=1-{\hat\gamma^2}_*(f),$ and (\ref{eq:cohdist}) holds.
\end{remark}

In a different context Gupta \cite{Gupta71} developed the distribution of the product of $p$ independent beta distributions:  a likelihood ratio criterion for testing a hypothesis about regression coefficients in a multivariate normal setting takes the form $\Lambda=\det\{\mbv_1\}/\det\{\mbv_1+\mbv_2\}$  under  the corresponding null hypothesis, with $\mbv_1$ and $\mbv_2$  independently distributed as 
$$
\mbv_1\EqualDist {\cal W}^{C}_{p}(f_1, {\bSigma}),\,\,
\mbv_2\EqualDist {\cal W}^{C}_{p}(f_2, {\bSigma}),
$$
for integer parameters $f_1, f_2$ and covariance matrix $\bSigma.$ Then
$\Lambda$ has the three-parameter complex $U$ distribution $U(p,f_2,f_1)$ which is distributed as a product of $p$ beta variables with $B_j {\displaystyle\EqualDist}\,\, {\rm beta}(f_1-j+1,f_2).$ So 
setting Gupta's parameters $f_1$ and $f_2$ to $K-p$ and $p$, respectively,  shows that $T(f)$ has the three-parameter complex $U$ distribution $U(p,p,K-p).$
This helps only a little because there are no simple expressions for this distribution's PDF or quantiles etc. However, by using convolution techniques Gupta did obtain some exact results for the case $p=2.$ In fact it turns out that 
for $p=2$  the right-side of (\ref{eqn:BoxApprox}) can be improved to
\begin{equation}\label{eq:correctedtwo}
\frac{K}{K-2} G(1-\alpha) \chi^2_{8}(1-\alpha)
\end{equation}
where   $G(1-\alpha)$ is an exact (tabulated) correction factor
and $\chi_{8}^{2}(1-\alpha)$ is the $100(1-\alpha)\%$ point of the chi-square distribution with $8$ degrees of freedom.  For example for $p=2, K=6$ and $\alpha=(0.05, 0.01)$ the factors are $(1.043, 1.051)$
\cite[Table 1]{Gupta71}.
The work of Gupta was extended as part of \cite[p.~5]{Krishnaiah_etal83} who produced tables of approximate correction factors 
for the right-side of (\ref{eqn:BoxApprox}) for $p\geq 3$ so that $M(f)$ is compared to 
\begin{equation}\label{eq:corrected}
\frac{K}{K-p}G(1-\alpha)\chi_{2p^{2}}^{2}(1-\alpha).
\end{equation}
Setting their parameters $n$ and $q$ to $K-p$ and $p$ respectively,   shows that for example for $p=3, K=8$ and $\alpha=(0.05, 0.01)$ the factors are $(1.076, 1.087)$ \cite[Table 7]{Krishnaiah_etal83}. The effect of these correction factors will be discussed shortly.

\begin{remark}
The result (\ref{eq:prodbeta}) is very nice, and quantiles of $T(f)$ could be found through, say, successive convolution techniques, but this is very complicated --- see \cite{Carter_etal76,Gupta73} who develop this approach for a related statistic.
\end{remark}

\subsection{Matching the first three cumulants exactly}\label{subsec:Fapprox}
The look-up tables of \cite{Gupta71} and \cite{Krishnaiah_etal83} are not convenient and so we now develop a simple and fast method for approximating the percentage points of the distribution of $M(f).$ 
Box \cite{Box49} considered using the very flexible Pearson system  for approximating the distribution of likelihood ratios. Box \cite[p.~330]{Box49} introduced a discriminant $d=(\kappa_1 \kappa_3)/(2\kappa_2^2),$ 
such that if $d>1$ a Pearson type VI should be fitted; this corresponds to $W_2 > W_1^2.$ For $p=2:20, K=1:100,$ with $K\geq2p$ we always found $d>1$ using 
(\ref{eq:cums}).  (Note $p=1$ is excluded since $W_2=W_1^2$ in that case.)

Box \cite{Box49} considered distributions of the form $bF_{\nu_1,\nu_2},$ i.e., a scaled $F$ distribution (Pearson type VI) with parameters $\nu_1,\nu_2,$ and suggested matching cumulants {\it approximately}. 

We have chosen to match the first three cumulants of the form (\ref{eq:cums}) {\it exactly}\,; the parameters of  $bF_{\nu_1,\nu_2}$ are related to the cumulants via \cite{Ginzberg13}
\begin{eqnarray}\label{eq:Fparams}
b \!\!&=&\!\! \frac{2\kappa_1\left(\kappa_1^2\kappa_2-\kappa_2^2+\kappa_1\kappa_3\right)}
{2\kappa_1^2\kappa_2-4\kappa_2^2+3\kappa_1\kappa_3},\nonumber\\
\nu_1 \!\!&=&\!\!\frac{4\kappa_1\left(\kappa_1^2\kappa_2-\kappa_2^2+\kappa_1\kappa_3\right)}
{4\kappa_1\kappa_2^2-\kappa_1^2\kappa_3+\kappa_2\kappa_3},\\
\nu_2 \!\!&=&\!\! \frac{4\kappa_1^2\kappa_2-8\kappa_2^2+6\kappa_1\kappa_3}{\kappa_1\kappa_3-2\kappa_2^2}.
\nonumber\end{eqnarray}
Then to carry out the test $M(f)$ would be compared to
\begin{equation}\label{eq:Fmethod}
b F_{\nu_1,\nu_2}(1-\alpha),
\end{equation}
where $F_{\nu_1,\nu_2}(1-\alpha)$ is the $100(1-\alpha)\%$ point of the $F$ distribution with parameters $b, \nu_1, \nu_2$ given by (\ref{eq:Fparams}).

\subsection{Comparison of Approximations}
For some combinations of $(p, K)$ 
the asymptotic result (\ref{eqn:Wilks})
 is compared to Box's basic approximation (\ref{eqn:BoxApprox}), the adjusted Box method (\ref{eq:correctedtwo}), (\ref{eq:corrected}) and the scaled $F$ method (\ref{eq:Fmethod}) in Table~\ref{tab:percentiles} which gives the $95\%$ and $99\%$ points of the distribution of $M(f)$ according to the four approaches. There is very good agreement between the adjusted Box method and the scaled $F$ method, the latter being quick and simple to compute. Box's basic approximation is a massive improvement on the asymptotic result.  For $p=2$ the adjusted Box approximation due to \cite{Gupta71} is exact and we see that the scaled $F$ approximation is therefore very accurate. Other combinations of $p$ and small $K$ lead to similar results. The agreement of the scaled $F$ approximation with the previous historically tabulated results (adjusted Box approximation) leads us to the following recommendation.
\begin{table}
\begin{center}
\[\begin{array}{|c||c|c|c|}
\hline
(p,K)  &  {\rm Method}     &  \alpha=0.05   &\alpha=0.01  \\ \hline \hline
(2,6) & {\rm Asymptotic}  &  15.51  & 20.09 \\
   & {\rm Box}  &  23.26  & 30.14 \\
   & {\rm Adjusted~Box}     & 24.26   & 31.67 \\ 
   & {\rm scaled}\, F     & 24.26   & 31.68 \\ 
 \hline
(3,8) & {\rm Asymptotic}  & 28.87  & 34.81 \\
   & {\rm Box}  &  46.19  & 55.69 \\
   & {\rm Adjusted~Box}     & 49.70   & 60.53 \\ 
   & {\rm scaled}\, F     & 49.71   & 60.54 \\ 
\hline
(4,10) & {\rm Asymptotic}  &  46.19  & 53.49 \\
   & {\rm Box}  &  76.99  & 89.14 \\
   & {\rm Adjusted~Box}     & 84.84   & 99.31 \\ 
   & {\rm scaled}\, F     & 84.85   & 99.30 \\ 
\hline
(5,12) & {\rm Asymptotic}  &  67.50  & 76.15 \\
   & {\rm Box}  &  115.72  & 130.55 \\
   & {\rm Adjusted~Box}     & 129.96   & 148.17 \\ 
   & {\rm scaled}\, F     & 129.94   & 148.18 \\ 
\hline

\end{array}
\]
\end{center}
\caption{Comparison of Percentage Points of $M(f)$ according to the asymptotic result (\ref{eqn:Wilks}), Box's approximation (\ref{eqn:BoxApprox}), adjusted Box method (\ref{eq:correctedtwo}), (\ref{eq:corrected}) and the scaled $F$ method (\ref{eq:Fmethod}). 
\label{tab:percentiles}}
\end{table}

\subsection{Recommended testing approach}
In view of the discusssions and results above, the following is recommended for a given choice of $\alpha:$
\begin{itemize}
\item 
If $p=1,$ reject $H_{0}$  if
\begin{equation}\label{eqn:transfrule}
M(f) > \frac{K}{K-1}\chi_{2}^{2}(1-\alpha).
\end{equation}
This test is distributionally exact.
\item
If $p\geq 2,$ reject $H_{0}$  if
\begin{equation}\label{eqn:Frule}
M(f)> bF_{\nu_1,\nu_2}(1-\alpha).
\end{equation}
The accuracy of the scaled $F$ approximation for our time series test (\ref{eq:TSsetting}) is now confirmed by simulation.
\end{itemize}

\section{Simulation Results}\label{sec:SimulationT}

\begin{figure}[t]
\begin{center}
\includegraphics[height=2.4in,width=3.4in]{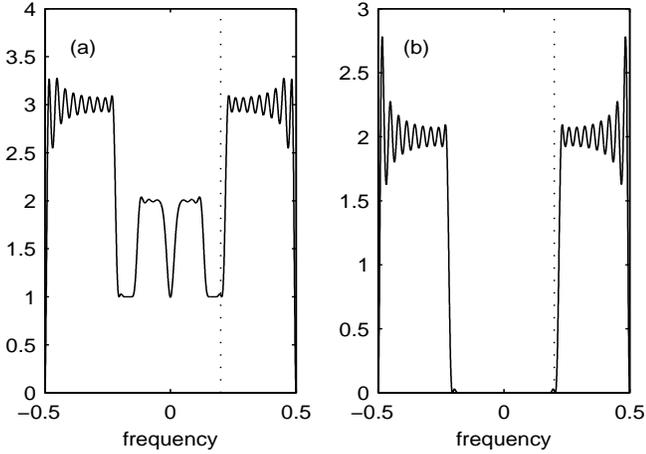}
\end{center}
\caption{\label{fig:SzRz}
(a) $S_Z(f)$ and (b) $R_Z(f).$ The vertical dotted line marks the frequency $f=0.2.$
}
\end{figure}
For $p\geq 2$ we will show that using the scaled  $F$ approximation test where we reject $H_0$ if (\ref{eqn:Frule}) holds brings about a worthwhile accuracy improvement over Box's approximation test
where we reject $H_0$ if
\begin{equation}\label{eqn:rule3}
M(f) > \frac{K}{K-p}\chi_{2p^2}^{2}(1-\alpha).
\end{equation}

To be able to do this
we need to simulate from a model such that
${\bS}_{\mbu}(f)$ in (\ref{eqn:suf}) has $\bR_{\mbZ}(f)={\bf 0}$ for some frequency range. We can proceed as follows.

We know \cite{PicBon97} that any complex second-order stationary scalar
process (assumed zero mean here), whether proper or improper,
can be written as the output of a widely linear filter driven by proper white noise, i.e.,
\begin{equation}\label{eq:zwl}
Z_t= \sum_{l=-\infty}^\infty g_l \epsilon_{t-l} + \sum_{l=-\infty}^\infty h_l \epsilon_{t-l}^*,
\end{equation}
where $\{ g_l \}$ and $\{ h_l \}$ are sequence of complex
constants, and $\{ \epsilon_{t}\}$ is proper white noise 
for which
$
\mbox{cov}\{ \epsilon_{t+\tau}, \epsilon_t \}
 =  \sigma_{\epsilon}^2 \delta_{\tau,0} \quad\mbox{and}\quad
\mbox{cov}\{ \epsilon_{t+\tau}, \epsilon_t^* \} = 0, 
$
for $\tau \in {\mathbb{Z}},$ where $\delta_{j,k}$ is the Kronecker delta.
For simulation purposes it is convenient to set
$\sigma_\epsilon^2=1.$  Then \cite{PicBon97} 
\begin{eqnarray}
S_Z(f)&=&  |G(f)|^2 + |H(f)|^2\label{eq:wl1}\\
R_Z(f)&=& G(f)H(-f)+ G(-f)H(f)\label{eq:wl2},
\end{eqnarray}
where $G(f)$ is the frequency response function of $\{ g_l \}$
given by $G(f)= \sum_{l=-\infty}^{\infty} g_l \re^{-\ri 2 \pi fl}$
and $H(f)$ is the frequency response function of $\{ h_l \}.$

For $p\geq 2$ we generate  processes $\{ Z_{j,t}\}, j=1,\ldots,p,$  such that
\begin{eqnarray}
Z_{j,t}&&= \sum_{l=-\infty}^\infty g_l \epsilon_{j,t-l} + \sum_{l=-\infty}^\infty h_l \epsilon_{j,t-l}^*\\
&&\qquad + \sum_{l=-\infty}^\infty a_l {\bar\epsilon}_{j,t-l} + \sum_{l=-\infty}^\infty a_l {\bar\epsilon}_{j,t-l}^*,
\end{eqnarray}
where the $2p$ processes $\{ \{\epsilon_{j,t}\},\{{\bar\epsilon}_{j,t}\}, j=1,\ldots,p\}$  are all independent of each other. The filter $\{ g_l \}$ was chosen to be low-pass with a frequency transition zone $[0.125, 0.15].$ The filter $\{ h_l \}$ was of `Hilbert-type' or all-pass in the frequency zone $[0.05, 0.45].$ Thus $G(f)$ is real and symmetric while $H(f)$ is imaginary and skew-symmetric.
According to (\ref{eq:wl2}), if using just these two filters, the resulting $R_Z(f)$ is zero for $f\in [-0.5,0.5].$ However, the filter $\{a_l\}$ was chosen to be high-pass above $f=0.2$ and therefore generates non-zero $R_Z(f)$ values at these high frequencies. The resulting $S_Z(f)$ and $R_Z(f)$ are shown in Fig.~\ref{fig:SzRz}.

The matrix $\bS_{\mbZ}(f)$ is thus of the form $\bS_{\mbZ}(f)= S_{\mbZ}(f)\bI_p$ with frequency dependence as shown in Fig.~\ref{fig:SzRz}(a) while $\bR_{\mbZ}(f)$ is of the form $\bR_{\mbZ}(f)= R_{\mbZ}(f)\bI_p$ with frequency dependence as shown in Fig.~\ref{fig:SzRz}(b). We can thus simulate from this model to evaluate our hypothesis tests, knowing that 
for frequencies where $\bR_{\mbZ}(f)={\bf 0}$ in fact $\bS_{\mbZ}(f)\not={\bf 0}$ and thus (\ref{eqn:glrformtwo}) is well-defined.

\begin{table}
\begin{center}
\[\begin{array}{|c||c|ccccc|}
\hline
 &  & &&f&&  \\ 
(p,K)  &  100\alpha\%     &   0.06    &0.12 & 0.18 & 0.24 & 0.42  \\ \hline \hline
(2,6) & 1\% &  1.5  & 1.5 & 1.4 \quad\vline& 31.1 & 35.4 \\
   &      & 1.1   & 1.1 & 0.9\quad\vline & 25.8 & 30.0 \\ 
 \cline{2-7}
& 5\% & 6.1   & 6.2 & 6.3 \quad\vline & 60.0 & 63.6  \\
&   & 5.0 & 5.1 & 5.2\quad\vline & 55.3 & 59.3 \\ 
\hline\hline
(3,8) & 1\% & 2.0   & 2.1 & 2.2 \quad\vline& 55.7 & 58.9\\
&  & { 0.9} & { 1.1} & { 1.1}\quad\vline & { 42.3} & { 45.5}  \\ 
 \cline{2-7}
& 5\% & 8.2   & 8.3 & 8.3 \quad\vline& 81.0 & 82.9  \\
&  & {4.9} & { 5.1} & { 5.2}\quad\vline & { 72.6} & { 75.2}  \\ 
 \hline
\end{array}
\]
\end{center}
\caption{Rejection percentages over 10\,000 repetitions. The top line of each entry is for
Box's  $\chi^2$ approximation (\ref{eqn:BoxApprox}) and the lower line
is for the $F$ approximation of (\ref{eqn:Frule}). 
\label{tab:one}}
\end{table}

Sample results are shown in Table~\ref{tab:one} for $(p,K)=(2,6)$ and $(3,8).$ So here $K=6$ and 8 are indeed small. 
Here $N=512$ but smaller time series lengths such as 128 produced very similar results.
Shown are 
rejection percentages for $H_0$ over $10\,000$ independent repetitions. The nominal rates are shown in the second column. The first three columns of rejection percentages are for frequencies where  $\bR_{\mbZ}(f)={\bf 0},$ ($H_0$ is true)
and the latter two are for frequencies where  $\bR_{\mbZ}(f)\not={\bf 0}$ ($H_0$ is false)
--- see  Fig.~\ref{fig:SzRz}(b). The top line of each entry is for
Box's  $\chi^2$ approximation (\ref{eqn:rule3}) and the lower line
is for the $F$ approximation of (\ref{eqn:Frule}). We see that, proportionately, the latter has a much more accurate rejection rate than Box's approximation when $H_0$ is true, but is slightly less accurate when $H_0$ is false.

\section{Data Analysis}\label{sec:realdata}
Here we apply our results to 
ocean current speed and direction time series recorded at
a mooring in the Labrador Sea \cite{ChandnaWalden11,Lillyetal99,LillyRhines02}. We
associate the eastward (zonal) measurement of current speed with
$\{ X_t \}$ and the northward (meridional) measurement with $\{ Y_t \}$
and thus obtain the complex-valued series from (\ref{eq:constructZ}). 
Series were recorded at six depths, (110, 760, 1260, 1760, 2510 and 3476m). 
The series are labelled 1 to 6 with increasing depth. We used $N=1600$ observations for the $6$-vector-valued complex time series, with a sampling interval of $\deltt=1$hr. In the spectral analysis $K=12$ sine tapers 
were applied.
Since $W_N$ in (\ref{eq:WNdef}) is $0.004$c/hr, the validity range $W_N \leq |f| \leq \fN-W_N$ for our statistical results
for a finite-$N$ sample is given by $0.004 \leq |f| \leq 0.496$c/hr. There was no evidence to reject 
the Gaussian assumption for this data set \cite{ChandnaWalden13}.

Of great interest to oceanographers are deep ocean motions well away from boundaries, especially in the internal wave frequency band.
We pay special attention to low frequencies $f\in[0.02, 0.14], $ in the internal wave band and near to the semi-diurnal tidal frequency.  The  so-called `inertial frequency' is approximately $0.07$c/hr for this latitude and purely clockwise rotation 
occurs at the inertial frequency in the Northern hemisphere, making a band centred around the inertial frequency particularly interesting to study for such complex-valued processes. The dominant semi-diurnal tide at around $f=0.08$c/hr was estimated and removed to avoid 
spectral leakage affecting estimation near the inertial frequency.

For this data $\bZ_t=[Z_{1,t},\ldots,Z_{6,t}]^T.$ In order to use different depth-contiguous sets of series we shall use the shorthand $\bZ_{m:m'} {\,{\displaystyle{\EqualDef }}\,} [Z_{m,t},\ldots,Z_{m',t}]^T$ with $1 \leq m<m'\leq 6.$
\subsection{Concentration of Canonical Coherencies}
The degree of polarization of a single random vector measures the spread amongst the eigenvalues of its covariance matrix. A random vector is completely polarized/unpolarized if all of its energy is concentrated in one direction/equally distributed amongst all dimensions.  This idea can be extended to the correlation between two random vectors by defining the correlation spread \cite{Schreier08} which provides a single, normalized 
measure of how much of the overall correlation is concentrated in a few coefficients, i.e., correlation is contained in a low dimensional subspace. 

Using the analogous definition to \cite{Schreier08} in our context we have 
\textit{coherence spread} defined by
\begin{equation}
\sigma^{2}_p(f){\,{\displaystyle{\EqualDef }}\,}\frac{p}{p-1}\left( \frac{\sum_{i=1}^p l^{4}_{j}(f)}{(\sum_{i=1}^p l^{2}_{j}(f) )^{2}}-\frac{1}{p}\right).
\end{equation}
If only one canonical coherence is non-zero, then $\sigma_p^{2}(f)=1,$ whereas if all canonical coherences are equal, $\sigma^{2}_p(f)=0.$ 
We note that if for  a given $f$, $\sigma^{2}_p(f)=1$, the likelihood ratio test statistic $T(f)=\prod_{j=1}^{p} (1-l^{2}_{j}(f))=0,$ i.e. achieves its minimum value, implying  ${\bR}_{\bZ}(f)\not={\bf 0}.$ Of course, for $\sigma_p^{2}(f)<1$ we are not able to conclude anything. In practice, 
we can only obtain an estimate $\hat{\sigma}_p^2(f)$ --- where the $l_j^2(f)$ are replaced by the $\ell_j^2(f)$ --- and therefore, the hypothesis test must be used to check for  ${\bR}_{\bZ}(f)={\bf 0}.$

\begin{figure}[t!]
\centering
\hspace*{-0.15in}
\includegraphics[height=3in,width=3.8in]{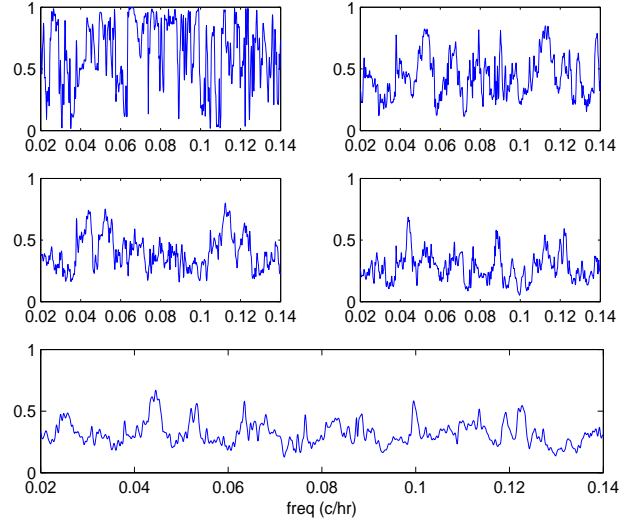}
\caption{Coherence spread estimate $\hat{\sigma}_p^2(f)$ for time series vectors (left to right, top to bottom): (a) $\bZ_{1:2}$, (b) $\bZ_{1:3}$, (c) $\bZ_{1:4},$ (d) $\bZ_{1:5}$ and (e) $\bZ_{1:6}$.
}\label{fig:coherencespread}
\end{figure}

Fig.~\ref{fig:coherencespread} displays $\hat{\sigma}_p^2(f)$ for vector time series  (a) $\bZ_{1:2}$, (b) $\bZ_{1:3}$, (c) $\bZ_{1:4}$, (d)$\bZ_{1:5}$ and (e) $\bZ_{1:6}$. An immediate observation is that the coherence spread estimate for $\bZ_{1:2}$ is highly erratic, with many values close to one. This is in contrast to all other plots where the spread ranges from $0.15-0.8$ gradually decreasing in range as we consider time series at increasing depths. A notable feature of (b) $\bZ_{1:3}$ is the broader peaks around $0.05, 0.065$ and $0.11$ and we see how the spread changes as we go from (b) $\bZ_{1:3}$ to (c) $\bZ_{1:4}$ with the broader peaks at $0.05$ and $0.11$ remaining intact whereas the one at $0.065$ shrinks from its value of $0.7$ to $0.4$; the sharper peaks at $0.08, 0.09$ and $0.138$ disappear and a new peak  appears at $0.044$ which persists in both (d) $\bZ_{1:5}$ and (e) $\bZ_{1:6}$. 
We have thus seen how an additional series (depth) notably changes the concentration level of the overall coherence at some frequencies while disturbing it much less at others.

\subsection{Test for Propriety}
As defined in Section~\ref{subsec:defproper} the process $\{ \bZ_t \}$ is proper when ${\bR}_{\bZ}(f)={\bf 0}\,\,\text{for all}\,\, |f|\leq \fN.$ 
Our test for
$
H_{0}: \bR_{\mbZ}(f)={\mathbf{0}}
$ 
is valid, and may be carried out, 
for any $W_{N}<f<\fN-W_{N}.$

\begin{figure}[t!]
\centering
\hspace*{-0.15in}
\includegraphics[height=3in,width=3.8in]{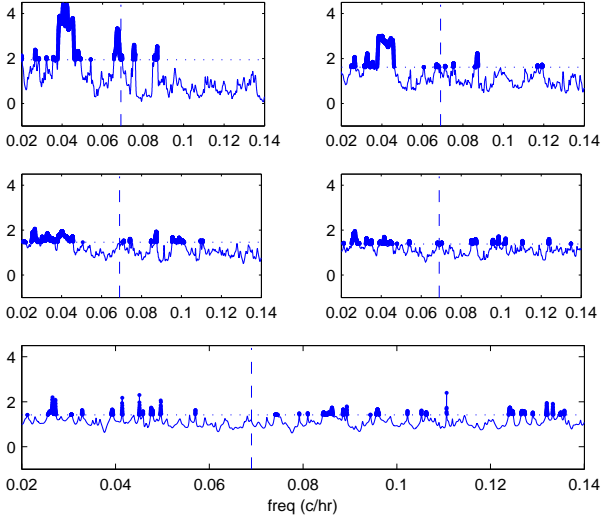}
\caption{The test statistic $M(f)$ (solid) and the critical value (dotted  line) for (left to right, top to bottom): (a) $\bZ_{1:2}$, (b) $\bZ_{1:3}$, (c) $\bZ_{1:4},$ (d) $\bZ_{1:5}$ and (e) $\bZ_{1:6}$. The test rejects the null hypothesis of propriety at frequencies where $M(f)$ exceeds the critical value (thick line portions). The dashed line is the semi-diurnal tidal frequency. 
}\label{fig:hypt_cum} 
\end{figure}

We test the same sets of  time series for propriety and the results are displayed in Fig.~\ref{fig:hypt_cum}. The solid line shows the test statistic $M(f)$ and the dotted line shows the critical value for each case. The test rejects $H_0$ at frequencies where $M(f)$ exceeds the critical value (thick line portions). The dashed line is the semi-diurnal tidal frequency. The coherence spread for $\bZ_{1:2}$ (first subplot in Fig.~\ref{fig:coherencespread}) takes the maximum value of $0.9927$ at $f=0.065,$ very close to the inertial frequency, and  Fig.~\ref{fig:hypt_cum} (a) shows that our test rejects $H_0$ around this frequency very clearly. The band of frequencies around $0.04$ is most prominent with rejection also clearly visible at frequencies $0.027, 0.075$ and $0.087$. For $\bZ_{1:3}$, the test rejects $H_0$ for almost the same set of low frequencies with  rejection also at a higher frequency around $0.12$. Results for $\bZ_{1:4}$ are very similar to that for $\bZ_{1:3}$, the main difference being that a small frequency band near $0.1$ also rejects $H_0.$ In general, we see that as other series (deeper in the ocean) are considered, $H_0$ is rejected, not only at low frequencies but also due to some additional higher frequencies,  but less definitively so. Importantly then, Fig.~\ref{fig:hypt_cum} shows {\it which}\/ frequency bands cause propriety to be rejected.

\section{Other Measures of Vector Coherence}\label{sec:othervc}
From Lemma~\ref{lem:one}, one measure for vector coherence is the sum of all the canonical coherencies:
\begin{equation}\label{eqn:secondstat}
\tr\{\bS_{\mbZ}^{-1}(f)\bR_{\mbZ}(f)\bS^{-T}_{\mbZ}(-f)
\bR_{\mbZ}^H(f)\}=\sum_{j=1}^p l_j^2(f).
\end{equation}

Levikov and Sokolov \cite{LevikovSokolov97} looked for a coefficient of coherence in the case of two  real-valued vector random processes. In our paper, for the vector 
$\bv_{t}=[\bX_{t}^{T},\bY^{T}_{t}]^{T},$ we consider $\{\bX_t\}$ and $\{\bY_t\}$ to be two geometrically related  vector components and combine them to form a complex-valued vector time series. Levikov and Sokolov did not consider the two processes to be  related  in such a way and treated them simply as two vector process. They did, however, make use of the frequency domain and derived the quantity
$$
\beta^2(f) \EqualDef \frac{1}{2} [ \bP(f) \bS_{\bY\bY}^{-1}(f)+
\bS_{\bY\bY}^{-1}(f) \bP(f)],
$$
where $\bP(f)=\bS_{\bY\bX}(f) \bS_{\bX\bX}^{-1}(f) \bS_{\bY\bX}^H(f).$ Taking the trace of this quantity we get
\begin{eqnarray}
\!\!\!\!\!\!\!
\tr\{\beta^2(f)\} \!\!\!\!&=& \!\!\!\!\frac{1}{2} \tr\{ \bP(f) \bS_{\bY\bY}^{-1}(f)+
\bS_{\bY\bY}^{-1}(f) \bP(f)\}\nonumber\\
\!\!\!\!&=& \!\!\!\! \tr\{  \bP(f) \bS_{\bY\bY}^{-1}(f) \}\nonumber\\
\!\!\!\!&=& \!\!\!\! \tr\{ \bS_{\bY\bX}(f) \bS_{\bX\bX}^{-1}(f) \bS_{\bY\bX}^H(f)\bS_{\bY\bY}^{-1}(f) \}\nonumber\\
\!\!\!\!&=& \!\!\!\! \tr\{ \bS_{\bX\bX}^{-1}(f) \bS_{\bX\bY}(f)\bS_{\bY\bY}^{-1}(f)\bS_{\bY\bX}(f)  \}.\label{eq:thirdstat}
\end{eqnarray}
This is of the same form as (\ref{eqn:secondstat}) only now using the components of the partition in (\ref{eqn:svf}); it will be the sum of all the canonical coherencies between $\dif \bZ_\bX(f)$ and $\dif \bZ_\bY(f)$ where
$$ \bX_t=\int_{-\fN}^{\fN} \e^{\eye 2\pi ft}\,\dif \bZ_\bX(f);\,\,
\bY_t=\int_{-\fN}^{\fN} \e^{\eye 2\pi ft}\,\dif \bZ_\bY(f).
$$

Now $\bZ_t=\bX_t+\eye \bY_t$ and $\bZ^*_t=\bX_t-\eye \bY_t.$ The spectral representation gives 
$\dif \bZ(f)=\dif\bZ_\bX(f)+\eye\, \dif\bZ_\bY(f)$ and $
\dif \bZ^*(-f)=\dif\bZ_\bX(f)-\eye\, \dif\bZ_\bY(f).
$
So we can write
\begin{equation}\label{eq:whattransform}
\left[
\begin{matrix}
 \dif \bZ(f)\\ 
\dif \bZ^*(-f) 
\end{matrix}
\right]
= {\bT} 
\left[
\begin{matrix}
 \dif \bZ_\bX(f)\\ 
\dif \bZ_\bY(f) 
\end{matrix}
\right],
\end{equation}
where $\bT$ is given in (\ref{eq:defT}). We know that affine transformations of $\dif \bZ_\bX(f)$ and of
$\dif \bZ_\bY(f)$ will not change  the canonical coherencies; however, (\ref{eq:whattransform}) does not represent affine transforms of $\dif \bZ_\bX(f)$ and of
$\dif \bZ_\bY(f)$ since a mixing is involved. 

Hence 
the quantities (\ref{eqn:secondstat}) and 
(\ref{eq:thirdstat}) will in general be different. Indeed for the example of
Section~\ref{sec:SimulationT} the value of (\ref{eq:thirdstat}) is zero over 
$|f|\leq 1/2,$ because $\bS_{\bX\bY}(f)=\bS_{\bY\bX}(f)={\bf 0}.$

\section{Summary and Conclusion}
We have developed a frequency domain approach to test for propriety of complex-valued vector time series. 
For propriety of $\{ \bZ_t \}$ we require ${\bR}_{\bZ}(f)={\bf 0}\,\,\text{for all}\,\, |f|\leq \fN.$ 
We can carry out the test
$
H_{0}: \bR_{\mbZ}(f)={\mathbf{0}}
$ 
for any $W_{N}<f<\fN-W_{N}.$
Most importantly for the vector case $(p\geq 2)$ we have justified use of the rule that $H_0$ is rejected if 
$
M(f) =-2K\log T(f)> bF_{\nu_1,\nu_2}(1-\alpha).
$
There is no assumption that $K$ is large, and indeed this would rarely be expected in practice. We have shown in detail how the  statistic $T(f)$  arises by consideration of canonical coherencies for complex-valued vector time series. When propriety is invalid, the frequency domain approach has the scientific advantage of showing which frequency bands are causing rejection, likely
allowing  linkage to known or hypothesized properties of the physical processes involved.

\appendix{}
\subsection{Proof of Lemma~\ref{lem:one} }\label{app:lem1}
Given (\ref{eq:standardize}), since $\bS_{\mbZ}(f)$ and $ \bS^{T}_{\mbZ}(-f)$ are positive-definite (Hermitian) covariance matrices,  we have solutions
\begin{eqnarray}
\bA^H(f) &=&\bF^H(f) \bS_{\mbZ}^{-1/2}(f)\label{eq:Aeq}\\
 \bB^H(f)&=&\bG^H(f) \bS^{-T/2}_{\mbZ}(-f)\label{eq:Beq}
\end{eqnarray}
where $\bF(f), \bG(f)\in {\mathbb C}^{p\times p}$ are unitary. Then
\begin{eqnarray}
\!\!\!\!\!\!\!\!\!\!\!\!\bK(f) &=&\bA^H(f)\bR_{\mbZ}(f)\bB(f) \nonumber\\
&=&
\bF^H(f) \bS_{\mbZ}^{-1/2}(f)\bR_{\mbZ}(f)\bS^{-T/2}_{\mbZ}(-f)\bG(f).
\label{eq:SVD}
\end{eqnarray}
We now make use of the weak majorization result  \cite[p.~294]{SchreierScharf10}. Let $|{\rm diag}(\bK(f))|\,{\displaystyle{\EqualDef}}\,[|K_{11}(f)|,\ldots,|K_{pp}(f)|]^T.$
Then
$$
\sum_{j=1}^r 
|K_{[jj]}(f)|\leq \sum_{j=1}^r |\sigma_{[j]}(f)|, \, r=1,\ldots,p.
$$
where the $\sigma_j(f)$ are the singular values  of $\bK(f)$ and  $\sigma_{[1]}(f)\geq \sigma_{[2]}(f)\geq \cdots \geq \sigma_{[p]}(f),$ (a descending size order).
Hence the solution to (\ref{eq:defineprob}) is found by making $\bK(f)$ diagonal. From  (\ref{eq:SVD}), we thus choose $\bF(f)$ and $\bG(f)$ to diagonalize $\bK(f),$ i.e.,
$\bF(f)$ and $\bG(f)$ are determined by singular value decomposition of 
$$
\bC(f)\EqualDef\bS_{\mbZ}^{-1/2}(f)\bR_{\mbZ}(f)\bS^{-T/2}_{\mbZ}(-f)=\bF(f) \bK(f)\bG^H(f),
$$
giving\begin{eqnarray*}
\bC(f)\bC^H(f)\!\!\!\! &=&\!\!\!\!\bS_{\mbZ}^{-1/2}(f)\bR_{\mbZ}(f)\bS^{-T}_{\mbZ}(-f)\bR^{H}_{\mbZ}(f)\bS_{\mbZ}^{-1/2}(f)\\
\!\!\!\! &=&\!\!\!\! \bF(f)\bL(f) \bF^H(f),
\end{eqnarray*}
where $\bL(f)$ denotes a diagonal matrix with $j$th element 
$l_j^2(f)=|K_{jj}(f)|^2,$ in descending size order. Now multiply through on the left by $\bS_{\mbZ}^{-1/2}(f)$ and on the right by $\bF(f)$ to obtain
\begin{eqnarray*}
&&\!\!\!\!\!\!\!\!\!\!\!\!\!\!\!\!\!\!\!\!\!\!\!\!\!\!\!\!\!\!\!\!
\bS_{\mbZ}^{-1}(f)\bR_{\mbZ}(f)\bS^{-T}_{\mbZ}(-f)\bR^{H}_{\mbZ}(f)\bS_{\mbZ}^{-1/2}(f)\bF(f)\\
\!\!\!\! &=&\!\!\!\! \bS_{\mbZ}^{-1/2}(f)\bF(f)\bL(f) ,
\end{eqnarray*}
which, using (\ref{eq:Aeq}), can be written
\begin{eqnarray*}
\bS_{\mbZ}^{-1}(f)\bR_{\mbZ}(f)\bS^{-T}_{\mbZ}(-f)\bR^{H}_{\mbZ}(f)\bA(f)=\bA(f)\bL(f) ,
\end{eqnarray*}
so that
$$
\bS_{\mbZ}^{-1}(f)\bR_{\mbZ}(f)\bS^{-T}_{\mbZ}(-f)\bR^{H}_{\mbZ}(f)\bA_j(f)=l_j^2(f) \bA_j(f),
$$
and $l_j^2(f)$ are the eigenvalues, and $\bA_j(f)$ are the eigenvectors 
of the $p\times p$ matrix $\bS_{\mbZ}^{-1}(f)\bR_{\mbZ}(f)\bS^{-T}_{\mbZ}(-f)\bR^{H}_{\mbZ}(f),$ as required. Note that this matrix is the product of the two Hermitian matrices $\bS_{\mbZ}^{-1}(f)$ and $\bR_{\mbZ}(f)\bS^{-T}_{\mbZ}(-f)\bR^{H}_{\mbZ}(f).$

Similarly,
\begin{eqnarray*}
\bC^H(f)\bC(f)\!\!\!\! &=&\!\!\!\!\bS^{-T/2}_{\mbZ}(-f)\bR^{H}_{\mbZ}(f)\bS_{\mbZ}^{-1}(f)\bR_{\mbZ}(f)\bS^{-T/2}_{\mbZ}(-f)\\
\!\!\!\! &=&\!\!\!\! \bG(f)\bL(f)\bG^H(f).
\end{eqnarray*}
Multiply through on the left by $\bS_{\mbZ}^{-T/2}(f)$ and on the right by $\bG(f),$ and use (\ref{eq:Beq}) to obtain
\begin{eqnarray*}
\bS_{\mbZ}^{-T}(-f)\bR_{\mbZ}^H(f)\bS^{-1}_{\mbZ}(f)\bR_{\mbZ}(f)\bB(f)=\bB(f)\bL(f),
\end{eqnarray*}
so that, as required
$$
\bS^{-T}_{\mbZ}(-f)\bR_{\mbZ}^H(f)\bS_{\mbZ}^{-1}(f)\bR_{\mbZ}(f)\bB_j(f)
=l_j^2(f) \bB_j(f).
$$
$l_j^2(f)$ are the eigenvalues, and $\bB_j(f)$ are the eigenvectors 
of the $p\times p$ matrix $\bS^{-T}_{\mbZ}(-f)\bR_{\mbZ}^H(f)\bS_{\mbZ}^{-1}(f)\bR_{\mbZ}(f).$ 

Notice that the matrices $\bS_{\mbZ}^{-1}(f)\bR_{\mbZ}(f)\bS^{-T}_{\mbZ}(-f)\bR^{H}_{\mbZ}(f)$ and $\bS^{-T}_{\mbZ}(-f)\bR_{\mbZ}^H(f)\bS_{\mbZ}^{-1}(f)\bR_{\mbZ}(f)$ are just cyclic permutations of each other.

Finally, 
$$
\cov\{ \dif \bZ_{\bxi}(f), \dif \bZ_{\beeta}(f)\}=E\{ \dif \bZ_{\bxi}(f)
\dif \bZ^H_{\beeta}(f)\}=\bK(f),
$$
and the solution of the optimization problem makes $\bK(f)$ diagonal.
Hence, ${\rm corr}\{\dif Z_{\xi_j}(f),\dif Z_{\eta_k}(f)\}=0, \,\,\,\mbox{for}\,\,\, j,k=1,\ldots,p; j\not=k.$

\subsection{Proof of Lemma~\ref{lemma:one}}\label{app:lemmaone}
To simplify notation we drop  explicit frequency dependence. 
Consider the distribution of $T=l_G^{1/K},$ given by
(\ref{eqn:formtouse}), {\it under the null hypothesis}. We have
$$
E\{ T^r\}=\int \cdots\int 
{\textstyle{\frac{[\det\{\bA\}]^r}{[\det\{\bA_{11}\}\det\{\bA_{22}\}]^r}}}
g({\bA}) \dif \bA,
$$
where $g(\bA)$ is the PDF for the complex Wishart distribution. As explained 
in the text under the null hypothesis we can take $\bS_{\mbu}$ to be $\bI_{2p}$ because of invariance under the group action. We can thus replace 
(\ref{eqn:ACWishart}) by
$$
\bA \EqualDist {\cal W}^{C}_{2p}(K, \bI_{2p}),
$$
and using \cite{Goodman63} we know that for $\bA>0, K\geq 2p,$
\begin{equation}\label{eq:defg}
g(\bA; K, 2p, \bI_{2p}) = c(K,2p)[\det\{\bA\}]^{K-2p} \e^{-\tr\{\bA\}},
\end{equation}
where $c(K,2p)$ is a constant defined by
\begin{equation}\label{eq:defcs}
c^{-1}(K,2p)=\pi^{p(2p-1)}\prod_{i=1}^{2p} \Gamma(K+1-i).
\end{equation}
So $E\{ T^r\}$ takes the form
\begin{eqnarray*}
&&  \!\!\!\!\!\!\!\!\!\!\!\!\!\!\!\!
c(K,2p) \int \cdots\int 
{\textstyle{\frac{[\det\{\bA\}]^{K-2p+r}}{[\det\{\bA_{11}\}\det\{\bA_{22}\}]^r}}}\e^{-\tr\{\bA\}} \dif \bA\\
&&\!\!\!\!\!\!\!\!\!\!\!\!\!\!\!\!
={{\frac{c(K,2p)}{c(K+r,2p)}}} \int \cdots\int 
{\textstyle{\frac{1}{[\det\{\bA_{11}\}\det\{\bA_{22}\}]^r}}}\\
&&\times\left[c(K+r,2p)[\det\{\bA\}]^{K-2p+r}\e^{-\tr\{\bA\}}\right] \dif \bA.
\end{eqnarray*}
The integration is w.r.t. $\dif\bA=\dif A_{11},\ldots,\dif A_{2p2p}.$ The term in the square brackets above is the PDF for the 
${\cal W}^{C}_{2p}(K+r, \bI_{2p})$ distribution. The integral of this density with respect to the elements in $\bA_{12}$ and $\bA_{21}$ must give the marginal density of $\bA_{11}, \bA_{22},$ which is the product 
\begin{equation}\label{eq:prodg}
g(\bA_{11};K+r,p,\bI_p) \cdot g(\bA_{22};K+r,p,\bI_p),
\end{equation}
since $\bA_{11}$ and $\bA_{22}$ are independent under the null hypothesis. Carrying out the integration and using (\ref{eq:defg}) and (\ref{eq:prodg}) we obtain
\begin{eqnarray}
&&\!\!\!\!\!\!\!\!\!\!\!\!\!\!\!
{{\frac{c(K,2p)}{c(K+r,2p)}}} \int \cdots\int 
{\textstyle{\frac{1}{[\det\{\bA_{11}\}\det\{\bA_{22}\}]^r}}}\nonumber\\
&&\!\!\!\!\!\!\!\!\!\!\!
\times\prod_{j=1}^2 c(K+r,p)[\det\{\bA_{jj}\}]^{K+r-p}{\e}^{-\tr\{\bA_{jj}\}} {\dif} {\bA}_{jj}\nonumber\\
&&\!\!\!\!\!\!\!\!\!\!\!\!
={{\frac{c(K,2p)}{c(K+r,2p)}}}\prod_{j=1}^2 \int \cdots\int c(K+r,p)[\det\{\bA_{jj}\}]^{K-p}
\nonumber\\
&&
\times\quad{\e}^{-\tr\{\bA_{jj}\}} {\dif} {\bA}_{jj}
\nonumber\\
&&={{\frac{c(K,2p)}{c(K+r,2p)}}}\prod_{j=1}^2
\frac{c(K+r,p)}{c(K,p)}.\label{eq:allthecs}
\end{eqnarray}
Using (\ref{eq:defcs}) and (\ref{eq:allthecs}) we get
\begin{equation}\label{eq:ETr}
\!E\{ T^r\}\!=\!\frac{\prod_{j=1}^{2p} \Gamma(K\!+\!r+1-j)
[\prod_{j=1}^p \Gamma(K\!+\!1-j)]^2}
{\prod_{j=1}^{2p} \Gamma(K\!+\!1-j)
[\prod_{j=1}^p \Gamma(K\!+\!r+1-j)]^2}.
\end{equation}
This agrees with \cite[eqn.~(2.6)]{Krishnaiah_etal76} which appears without reference or proof.
Now $T^r=l_G^{r/K}$ so if we let $r\rightarrow rK,$ then $T^{rK}=l_G^r.$ So
\begin{eqnarray*}
&&\!\!\!\!\!\!\!\!\!\!\!
E\{ l_G^r\}\!=\!\frac{\prod_{j=1}^{2p} \Gamma(K[1+r]+1-j)
\prod_{j=1}^p \Gamma(K\!+\!1\!-\!j)}
{\prod_{j=1}^{2p} \Gamma(K\!+\!1\!-\!j)
\prod_{j=1}^p \Gamma(K[1+r]+1-j)}\\
&&\times \left[ 
\frac{
\prod_{j=1}^p \Gamma(K\!+\!1\!-\!j)}
{
\prod_{j=1}^p \Gamma(K[1+r]+1-j)}
\right]\\
&&=\frac{\prod_{j=1}^{p} \Gamma(K[1+r]+1-j-p)
}
{
\prod_{j=1}^p \Gamma(K+1-j-p)}\\
&&\times
\left[ 
\frac{
\prod_{j=1}^p \Gamma(K\!+\!1\!-\!j)}
{
\prod_{j=1}^p \Gamma(K[1+r]+1-j)}
\right]\\
&&=\left[ 
\frac{
\prod_{j=1}^p \Gamma(K\!+\!1\!-\!j)}
{
\prod_{j=1}^p \Gamma(K+1-j-p)}
\right]\\
&&\times\frac{\prod_{j=1}^{p} \Gamma(K[1+r]+1-j-p)
}
{
\prod_{j=1}^p \Gamma(K[1+r]+1-j)},
\end{eqnarray*}
which is (\ref{eqn:moment}).

\subsection{Proof of Lemma~\ref{lemma:two}}\label{app:lemmatwo}
Under the null hypothesis the $r$th moment of $T(f)$ is
\begin{equation}\label{eq:firstbeta}
E\{ T^r(f)\}=\prod_{j=1}^p 
\frac{  \Gamma(K+r+1-j-p)\Gamma(K+1-j)  }
{\Gamma(K+r+1-j)\Gamma(K+1-j-p)}.
\end{equation}
To see this
start with (\ref{eq:ETr}) and proceeed in analogous vein to the last part of the proof of Lemma~\ref{lemma:one}; since we are continuing to look at $E\{ T^r \}$ the step $r \rightarrow rK$ is not made. Note that when $j=p$ the critical gamma function argument is still positive: $K+r+1-j-p=K+r+1-2p>0$ since $K\geq 2p$ with $r\geq 0.$

A real scalar random variable $X$ is said to have a (type-1) beta distribution, $X \,\,{\displaystyle\EqualDist}\,\, {\rm beta}(\alpha,\beta),$ 
if the PDF is 
$$
f(x)=\frac{\Gamma(\alpha+\beta)}{\Gamma(\alpha)\Gamma(\beta)} x^{\alpha-1}(1-x)^{\beta-1}, \,\,0<x<1, \alpha>0,\beta>0.
$$
The $r$th moment for this distribution is
\begin{equation}\label{eq:secondbeta}
E\{ X^r \}= \frac{\Gamma(\alpha+r)\Gamma(\alpha+\beta)}
{\Gamma(\alpha+\beta+r)\Gamma(\alpha)},\,\,\alpha+r >0.
\end{equation}
Comparing (\ref{eq:firstbeta}) and (\ref{eq:secondbeta}) we see for a fixed $j$ that $\alpha=K+1-j-p$ and  $\beta=p$ which gives the required result.

\section*{Acknowledgement}
The authors are  grateful to Jon Lilly for the Labrador Sea data.

\end{document}